\documentclass[11pt]{article}
\usepackage{path,graphicx}
\usepackage{a4,amssymb, dsfont}
\usepackage{authblk}
\usepackage{amsmath}
\usepackage{cite}
\usepackage[hyphens]{url}
\usepackage{stackrel}
\usepackage[margin=1.1in]{geometry}
\usepackage[utf8]{inputenc}
\usepackage{float}
\usepackage[utf8]{inputenc}
\usepackage{fancyhdr}
\newtheorem{definition}{Definition}[section]
\newtheorem{lemma}[definition]{Lemma}
\newtheorem{theorem}[definition]{Theorem}
\newtheorem{proposition}[definition]{Proposition}

\newcommand{\proof}{\noindent {\bf Proof. }}
\newcommand{\qed}{\hfill $\square$\smallskip}
\newtheorem{innercustomthm}{Theorem}
\newenvironment{customthm}[1]
{\renewcommand\theinnercustomthm{#1}\innercustomthm}
{\endinnercustomthm}

\newtheorem{innercustomp}{Proposition}
\newenvironment{customp}[1]
{\renewcommand\theinnercustomp{#1}\innercustomp}
{\endinnercustomp}
\title{Majority Opinion Diffusion in Social Networks: An Adversarial Approach}
\author{Ahad N. Zehmakan\thanks{Corresponding author; Email Address: abdolahad.noori@inf.ethz.ch, Postal Address: CAB G F 79, Institute of Theoretical Computer Science, ETH Z\"urich, Universit\"atstrasse 6, CH-8092 Z\"urich.}}
\affil{Department of Computer Science, ETH Zurich}
\date{} 
\begin{document}
\maketitle
\begin{abstract}
We introduce and study a novel majority-based opinion diffusion model. Consider a graph $G$, which represents a social network. Assume that initially a subset of nodes, called seed nodes or early adopters, are colored either black or white, which correspond to positive or negative opinion regarding a consumer product or a technological innovation. Then, in each round an uncolored node, which is adjacent to at least one colored node, chooses the most frequent color among its neighbors.
	
Consider a marketing campaign which advertises a product of poor quality and its ultimate goal is that more than half of the population believe in the quality of the product at the end of the opinion diffusion process. We focus on three types of attackers which can select the seed nodes in a deterministic or random fashion and manipulate almost half of them to adopt a positive opinion toward the product (that is, to choose black color). We say that an attacker succeeds if a majority of nodes are black at the end of the process. Our main purpose is to characterize classes of graphs where an attacker cannot succeed. In particular, we prove that if the maximum degree of the underlying graph is not too large or if it has strong expansion properties, then it is fairly resilient to such attacks.
	
Furthermore, we prove tight bounds on the stabilization time of the process (that is, the number of rounds it needs to end) in both settings of choosing the seed nodes deterministically and randomly. We also provide several hardness results for some optimization problems regarding stabilization time and choice of seed nodes.
\end{abstract}
\section{Introduction}
\noindent In real life, we usually have specific perspectives on various topics, such as consumer products, technological innovations, life styles, and political events and by communicating with friends, family, and colleagues, our opinions are influenced. Opinion diffusion and (mis)-information spreading can affect different aspects of our lives such as economy, defense, fashion, even personal affairs. Therefore, there has been a growing interest to understand how opinions form and diffuse because of the existence of social ties among a community's members and how the structure of a social network can influence this process. This would enable us to obtain better predictions of electoral results, control the effect of marketing and political campaigns, and in general advance our knowledge of the cognitive processes behind social influence. 

The study of opinion diffusion and the evolution of social dynamics on networks has attracted the attention of researchers from a vast spectrum of disciplines such as economics~\cite{bharathi2007competitive}, epidemiology~\cite{pastor2001epidemic}, social psychology~\cite{yin2019agent}, statistical physics~\cite{gartner2020threshold}, and political sciences~\cite{n2020rumor}. It has also gained significant popularity in theoretical computer science, especially in the quickly growing literature focusing on the interface between social choice and social networks, cf.~\cite{bredereck2017manipulating} and~\cite{auletta2018reasoning}.

From a theoretical viewpoint, it is natural to introduce and study mathematical models which mimic different opinion dynamics. Of course in the real world, they are too complex to be explained in purely mathematical terms. However, the main idea is to comprehend their general principles and make crude approximations at discovering certain essential aspects of them which are otherwise totally hidden by the complexity of the full phenomenon. 

In these models, it is usually assumed that we have a graph $G$ and initially some nodes are colored, say black or white. Then, in each round a group of nodes get colored or update their color based on a predefined rule. Graph $G$ is meant to represent a social network, where each agent is modeled as a node and edges indicate relations between them, e.g., friendship, common interests, advice, or various forms of interactions. Furthermore, black and white stand for the opinion of an agent regarding an innovation or a political party, etc. 

In the plethora of opinion diffusion models, threshold-based ones are certainly the best known, cf.~\cite{kempe2003maximizing},~\cite{apt2014social},~\cite{jeger2019dynamic}, and~\cite{zehmakan2020opinion}. There, nodes (i.e., agents) adopt a color (i.e., opinion) if it is shared by a certain number or fraction of their connections. Particularly, the majority-based models, where each node chooses the most frequent color among its neighbors, have received a substantial amount of attention, cf.~\cite{chistikov2020convergence}. This imitating behavior can be explained in several ways: an agent that sees a majority agreeing on an opinion might think that her neighbors have access to some information unknown to her and hence they have made the better choice; also agents can directly benefit from adopting the same behavior as their friends (e.g., prices going down).

Nowadays, the identification of factors leading to a successful innovation in a given market is a question of considerable practical importance. Marketing campaigns routinely use online social networks to attempt to sway people's opinions in their favor, for instance by targeting segments of agents with free sample of their products or misleading information. Consequently, the study of control and manipulation of collective decision-making has gained increasing popularity in mechanism design, algorithmic game theory, and computational social choice, cf.~\cite{bredereck2017manipulating}. Especially, majority-based models have been postulated as one potential explanation for the success or failure of collective action and the diffusion of innovations. For example for different majority-based opinion diffusion models, a substantial amount of attention has been devoted to the study of characterizing graph structures for which the dominant color at the end of the process is the same as the dominant color in the initial coloring, cf.~\cite{auletta2015minority} and~\cite{zehmakan2020spread}.

\subsection{Our Contribution}
We introduce a novel majority-based opinion diffusion model and consider three different types of attackers whose goal is to engineer the output of the diffusion process. Our central problem is to characterize classes of graphs for which an attacker fails to reach its goal. 

\paragraph{Our model.} Consider a graph $G$. Assume that initially a subset of nodes, which are called the \emph{seed nodes}, are colored black or white and the rest of nodes are uncolored. Then, in discrete-time rounds each node, which is uncolored and is adjacent to at least one colored node, chooses the most frequent color among its colored neighbors. In case of a tie, it chooses black color with probability (w.p.) $1/2$ independently and white otherwise.\footnote{We should mention that the results provided in the present paper actually hold for any choice of the tie-breaking rule.}

The seed nodes correspond to early adopters, who are the first customers to adopt a new product or technology before the rest of the population does and usually constitute 10-20\% of the population. They are often called lighthouse customers because they serve as a beacon of light for the rest of the population to follow, which will take the technology or product mainstream. The term early adopters comes from the technology adoption curve, which was popularized by the book Diffusion of Innovations~\cite{rogers1962diffusion}.

The seed nodes and their color might be chosen randomly or deterministically. We take an adversarial perspective and focus on the following three types of attackers, where we assume that $\alpha, \epsilon \in (0,1/2)$ are some arbitrary constants and $n$ is the number of nodes in the underlying graph.
\begin{definition}[strong attacker]
	An ($\alpha$,$\epsilon$)-strong attacker selects a seed set of size $\alpha n$ and color $(1/2+\epsilon)$ fraction of its nodes white and the rest black.
\end{definition}
\begin{definition}[moderate attacker]
	An ($\alpha$,$\epsilon$)-moderate attacker selects a seed set of size $\alpha n$ and color each seed node white, independently, w.p. $(1/2+\epsilon)$ and black otherwise.
\end{definition}
\begin{definition}[weak attacker]
	An ($\alpha$,$\epsilon$)-weak attacker selects each node to be a seed node, independently, w.p. $\alpha$ and then color each seed node white, independently, w.p. $(1/2+\epsilon)$ and black otherwise.
\end{definition}
Assume that the process runs for $t$ rounds. Then, the main goal of an attacker is to maximize the ratio of the number of black nodes to white ones at the end of the $t$-th round. We say that the attacker \emph{wins} if this ratio is at least one half. We are interested in bounding the probability that an attacker wins.

To better understand the aforementioned attacker models, you might think of an attacker as a marketer which desires to advertise an innovation or a product of poor quality and it can select the set of early adopters (i.e., seed nodes) in a random or deterministic fashion. Then, it manages to convince almost half of them to adopt a positive opinion about the product (i.e., choose black color). More precisely, a strong attacker can choose $(1/2-\epsilon)$ fraction of the seed nodes to be black and a moderate/weak attacker colors each seed node black w.p. $(1/2-\epsilon)$. The attacker's ultimate goal is that a majority of nodes are colored black after some number of rounds, even though initially white is the dominant color.

As an extreme example, consider a star graph $S_n$, which includes $n-1$ leaves and an internal node of degree $n-1$. An ($\alpha$,$\epsilon$)-strong attacker can choose the internal node and $\alpha n-1$ of the leaves to be the seed nodes and color the internal node black. Then after one round of the process, all the remaining $(1-\alpha)n$ nodes will be colored black. Thus, the attacker wins. Even for an ($\alpha$,$\epsilon$)-weak attacker (which essentially has no real selection power and only runs a random procedure), the internal node is selected w.p. $\alpha$ and will be colored black w.p. $(1/2-\epsilon)$. Hence, the attacker wins w.p. $(1/2-\epsilon)\alpha$.

Our main purpose is to characterize classes of graphs which are \textit{resilient} to such attacks (i.e., an attacker cannot win) where we assume that the attacker has full knowledge of the graph structure. We should emphasize that for a moderate or weak attacker since each seed node is white independently w.p. $(1/2+\epsilon)$, the probability that the attacker wins is at most $1/2$. However, we are interested in graph structures where the probability of winning is extremely small.

\paragraph{Strong attacker.} An ($\alpha$,$\epsilon$)-strong attacker is quite powerful. However, there are graphs, such as a complete graph, where it cannot win. We prove that if the graph is regular and has strong expansion properties, then a strong attacker fails. More precisely we prove that for a regular graph $G$ if $\sigma (G) \le \epsilon\sqrt{\alpha (1-\alpha)}$, then an ($\alpha$,$\epsilon$)-strong attacker cannot win, where $\sigma(G)$ is the second-largest absolute eigenvalue of the normalized adjacency matrix of $G$. (Please see Section~\ref{strong attacker} for a more detailed definition of $\sigma(G)$ and its relation to expansion.) We argue that expansion and regularity are not only sufficient conditions, but also somewhat necessary for a graph to be resilient to a strong attacker.

In a nutshell, since a strong attacker has the power to choose the seed nodes and their color, for a graph to be resilient, the ``influencing power'' should be distributed uniformly among all nodes. In particular, the number of edges among each two node sets must be proportional to their size and this is what basically regularity and expansion provide.

\paragraph{Moderate attacker.} Roughly speaking, if there is a small set of nodes of size $s$ with significant influencing power, then a weak attacker can select them to be in the seed set. These nodes will be colored black w.p. $(1/2-\epsilon)^s$, which is non-negligible for small $s$, and this can result in the attacker winning. A natural way to avoid such scenarios is to bound the maximum degree of the underlying graph. We show that if the maximum degree is not ``too large'', then a moderate attacker fails asymptotically almost surely. (We say an event happens asymptotically almost surely (a.a.s.) if it occurs with a probability tending to 1 while we let $n$ go to infinity.) More precisely, we prove that for a graph $G$ and an ($\alpha$,$\epsilon$)-moderate attacker, if $\Delta\le (Cn/\log (1/\mu))^{\frac{1}{2t}}$ for some small constant $C_{\epsilon,\alpha}>0$, the attacker cannot win in $t$ rounds w.p. at least $1-\mu$, where $\Delta$ denotes the maximum degree. Furthermore, we argue the tightness of this statement.

\paragraph{Weak attacker.} As we discussed, an ($\alpha$,$\epsilon$)-weak attacker wins on a star graph $S_n$ w.p. at least $(1/2-\epsilon)\alpha$. Loosely speaking, the graphs of this type, where there is a small subset of nodes with significant influencing power and a large set of nodes with very limited power, are problematic. Our goal is basically to show that graphs which do not fall under the umbrella of this type of structure are resilient to a weak attacker. We prove that if the maximum degree in a graph $G$ is sufficiently small or if a majority of nodes have rather large degrees, then a weak attacker will fail. More accurately, we show that for a graph $G$ and an ($\alpha$,$\epsilon$)-weak attacker, if 
$\Delta\le Cn/((\log n)^{C^{\prime}}\log (1/\mu))$
for some suitable constants $C_{\alpha, \epsilon}, C_{\alpha,\epsilon}^{\prime}>0$ or if half of nodes are of degree at least $C^{\prime\prime}\log (1/\mu)$ for a sufficiently large constant $C^{\prime\prime}_{\alpha,\epsilon}$, then the attacker fails w.p. at least $1-\mu$. Note that the bound on $\Delta$ does not depend on $t$ unlike the case of a moderate attacker.

\paragraph{Stabilization time.}
For a connected graph $G$, if the seed set is non-empty, then all nodes will be colored eventually. The number of rounds the process needs to color all nodes is called the \emph{stabilization time} of the process. As we discuss in Section~\ref{stabilization time}, it is straightforward to prove that the stabilization time is upper-bounded by the diameter of $G$. This is, in particular, true for the setting of a strong/moderate attacker. However, what if the seed nodes are chosen at random (for example, in case of a weak attacker)? We prove that if each node is in the seed set independently w.p. $\alpha$, then the stabilization time is a.a.s. lower-bounded by $\Omega(\log_{\Delta}\log n^{\frac{1}{\alpha}})$ and upper-bounded by $\mathcal{O}((1/\alpha\delta)\log n)$, where $\delta$ denotes the minimum degree in $G$. Furthermore, we argue that these bounds are both tight, up to a constant factor.

\paragraph{Hardness results.} We provide some hardness results for a decision problem on the stabilization time and an optimization problem on the choice of seed set and its coloring. Assume that we are given a graph $G$ and some integers $t$ and $s$. We prove that the problem of determining whether there is a seed set of size $s$ for which the process takes $t$ rounds to end is NP-complete. Furthermore, suppose we are given a graph $G$ and some integers $b$, $w$, and $t$ as the input and our goal is to find the minimum (expected) number of white nodes after $t$ rounds if there are $b$ black and $w$ white nodes initially. We prove the best possible approximation factor for this problem is larger than $n^{1-\zeta}$ for any $\zeta>0$ unless P=NP.

\subsection{Related Work}
Numerous opinion diffusion models have been introduced to investigate how a group of agents modify their opinions under the influence of other agents. It is usually assumed that for a graph $G$, which represents a social network, initially some nodes are colored black (positive) or white (negative). Then, nodes update their color based on some predefined rule. Among these models, the threshold and majority model are perhaps the closest to ours. In both of them, initially each node is black or white. In the \emph{majority model}~\cite{peleg1997local}, in each round all nodes simultaneously update their color to the most frequent color in their neighborhood (and no update in case of a tie). In the \emph{threshold model}~\cite{kempe2003maximizing}, each node $v$ has a threshold value $t_v$ and it becomes black as soon as it has at least $t_v$ black neighbors. 

We should emphasize that different variants of these models have been considered; for example, asynchronous updating rule~\cite{auletta2015minority}, various tie-breaking rules~\cite{peleg1998size}, random threshold values~\cite{kempe2003maximizing}, and with a bias toward one of the colors~\cite{anagnostopoulos2020biased}. Even more complex models such as the ones considered in~\cite{ferraioli2017social} and~\cite{auletta2019consensus}, which follow an averaging-based updating rule, or the models in~\cite{brill2016pairwise},~\cite{meir2010convergence},~\cite{aulettaeffectiveness},~\cite{zehmakan2019two}, and~\cite{faliszewski2018opinion} can be seen as extensions of the majority model. 

Consider a marketer which advertises a new product. Assume that it can convince a subset of agents to adopt a positive opinion about its product, e.g., by giving them free samples of the product, and it aims to trigger a large cascade of further adoptions. Which agents should it target? A node set $A$ is a \emph{target set} if black color eventually takes over all (or half) of nodes once $A$ is fully black. (This is sometimes also known as dynamic monopoly or percolating set.) For both the majority and threshold model, the minimum size of a target set has been extensively studied on various classes of graphs such as lattice~\cite{gartner2017color}, Erd\H{o}s-R\'{e}nyi random graph~\cite{schoenebeck2018consensus}, random regular graphs~\cite{gartner2018majority}, power-law random graphs~\cite{amini2014bootstrap}, expander graphs~\cite{mossel2014majority}, and bipartite graphs~\cite{zehmakan2019tight}. Furthermore, Berger~\cite{berger2001dynamic} proved that there exist arbitrarily large graphs which have target sets of constant size under the majority model and it was shown in~\cite{auletta2018reasoning} that every $n$-node graph has a target set of size at most $n/2$ under the asynchronous variant.

Furthermore, it is known that the problem of finding the minimum size of a target set for a given graph $G$ is NP-hard for different variants of these models. For the majority model, Mishra, Radhakrishnan, and Sivasubramanian~\cite{mishra2002hardness} proved that this problem cannot be approximated within a factor of $\log \Delta\log\log \Delta$, unless P=NP; however, there is a polynomial-time $\log\Delta$-approximation algorithm. For the threshold model, Chen~\cite{chen2009approximability} proved that the problem is NP-hard even when all thresholds $t_v$ are 1 or 2 and $G$ is a bounded-degree bipartite graph. On the other hand, the problem is traceable for special classes of graphs such as trees. We study a similar optimization problem in our setting and provide inapproximability results.

Many other adversarial scenarios where an attacker aims to engineer the output of an opinion diffusion process have been investigated. For example, assume that an attacker can manipulate almost half of initially colored nodes to choose black color in a random or deterministic manner, similar to our attacker models. Which graph structures are resilient to such attacks, i.e., at the end of the process the white color is still the dominant color? For the majority model, it was proven in~\cite{mustafa2001majority} that a graph is resilient if it is a clique or very close to a clique. In~\cite{auletta2015minority} similar results were provided for the asynchronous setting. For our majority-based model, we prove that expansion, regularity, and maximum degree are chiefly responsible for resilience.

Other kinds of attackers with various manipulation powers, such as adding/deleting edges or changing the order of updates, have been considered, cf.~\cite{bredereck2017manipulating},~\cite{coro2019recommending}, and~\cite{wilder2017controlling}. Moreover, the complexity of the problem of finding an optimal strategy for different types of attackers has been studied, cf.~\cite{grandi2018complexity},~\cite{coro2019exploiting}, and~\cite{borodin2010threshold}. 

In the majority model, since the updating rule is deterministic, the process eventually reaches a cycle of colorings. The length of this cycle and the number of rounds the process needs to reach it are called the periodicity and stabilization time of the process, respectively. It was proven in~\cite{goles1980periodic} that periodicity is always one or two. Recently, it was shown in~\cite{chistikov2020convergence} and~\cite{zhuanglifting} that it is PSPACE-complete to decide whether the periodicity is one or not for a given coloring of a \emph{directed} graph. Regarding the stabilization time, Fogelman, Goles, and Weisbuch~\cite{fogelman1983transient} showed that it is bounded by $\mathcal{O}\left(n^2\right)$. It was proven in~\cite{frischknecht2013convergence} that this bound is tight, up to some poly-logarithmic factor. It was shown in~\cite{kaaser2015voting} that the problem of determining whether there exists a coloring for which the process takes at least $t$ rounds is NP-complete. As mentioned, we provide tight bounds and hardness results on the stabilization time of our diffusion process.

\subsection{Preliminaries}
\paragraph{Graph definitions.} Let $G=\left(V,E\right)$ be an $n$-node graph. For two nodes $v,u \in V$, we define $d\left(v,u\right)$ to be the length of a shortest path between $v,u$ in terms of the number of edges, which is called the \emph{distance} between $v$ and $u$ (for a node $v$, we define $d\left(v,v\right):=0$). For $t\in \mathbb{N}_0$, we let $N_t(v):=\{u\in V: d(v,u)=t\}$ denote the set of nodes whose distance from $v$ is exactly $t$. In particular, $N_0(v)=v$ and $N_{1}(v)$ is the set of $v$'s \emph{neighbors}. Furthermore, we define $\hat{N}_t(v):=\cup_{i=0}^{t}N_i(v)$ to be the \emph{$t$-neighborhood} of $v$. Analogously, for a node set $S\subseteq V$ we have $N_t(S):=\{u\in V: d(S,u)=t\}$ and $\hat{N}_t(S):=\cup_{i=0}^{t}N_i(S)$, where $d(S,u):=\min_{v\in S}d(v,u)$. 
We let $\deg\left(v\right):=|N_1\left(v\right)|$ denote the \emph{degree} of $v$ and define $\deg_S\left(v\right):=|N_1\left(v\right)\cap S|$. We also define $\Delta\left(G\right)$ and $\delta\left(G\right)$ to be respectively the maximum and minimum degree in graph $G$. 
For two node sets $S$ and $S'$, we define $e\left(S,S'\right):=|\{\left(v,u\right)\in S\times S': \{v,u\}\in E\}|$
where $S\times S'$ is the Cartesian product of $S$ and $S'$. 

\paragraph{Model definitions.} In our model, we will denote by $R_t$ the set of nodes which are colored in the $t$-th round for $t\in \mathbb{N}_0$. In other words, $R_0$ is equal to the seed set and $R_t:=N_t(R_0)$ for $t\in \mathbb{N}$. We also define $\hat{R}_t=\cup_{i=0}^{t}R_i$. Moreover, we let $B_t$ (analogously $W_t$) denote the set of black (resp. white) nodes in $R_t$. Similarly, $\hat{B}_t$ and $\hat{W}_t$ denote the set of black and white nodes in $\hat{R}_t$, respectively. We also define $r_t:=|R_t|$, $\hat{r}_t=|\hat{R}_t|$, $b_t:=|B_t|$, $\hat{b}_t=|\hat{B}_t|$, $w_t=|W_t|$, and $\hat{w}_t=|\hat{W}_t|$. Note that these are random variables even when we choose the seed nodes and their color deterministically since we break a tie at random.

\paragraph{Some inequalities.} Now, we present some standard probabilistic inequalities, cf.~\cite{dubhashi2009concentration}, which we utilize several times later.

\begin{theorem}[Chernoff bound]
\label{Chernoff}
Suppose $x_1,\cdots,x_k$ are independent Bernoulli random variables and let $X$ denote their sum, then for any $0\leq \eta\leq 1$
\begin{itemize}
\item $\Pr[\left(1+\eta\right)\mathbb{E}[X]\leq X]\leq \exp\left({-\frac{\eta^2\mathbb{E}[X]}{3}}\right)$
\item$\Pr[X\leq \left(1-\eta\right)\mathbb{E}[X]]\leq \exp\left({-\frac{\eta^2\mathbb{E}[X]}{2}}\right)$.
\end{itemize}
\end{theorem}
If we are given $k$ discrete probability spaces $\left(\Omega_i,\Pr_i\right)$ for $1 \leq i \leq k$, then their product is defined to be the probability space over the ground set $\Omega:=\Omega_1 \times \Omega_2 \times \dots \times \Omega_k$ with the probability function $\Pr[\left(\omega_1, \dots, \omega_k\right)]:=\prod_{i=1}^{k}\Pr_i[\omega_i]$,
where $\omega_i \in \Omega_i$. 
Let $\left(\Omega,\Pr\right)$ be the product of $k$ discrete probability spaces, and let $X:\Omega\rightarrow \mathbb{R}$ be a random variable over $\Omega$. We say that the \emph{effect} of the $i$-th coordinate is at most $c_i$ if for all $\omega, \omega' \in \Omega$ which differ only in the $i$-th coordinate we have $|X\left(\omega\right)-X\left(\omega'\right)|\leq c_i$. Azuma's inequality states that $X$ is sharply concentrated around its expectation if the effect of the individual coordinates is not too big.
\begin{theorem}[Azuma's inequality]
\label{Azuma}
\label{theorem 6} Let $\left(\Omega,\Pr\right)$ be the product of $k$ discrete probability spaces $\left(\Omega_i,\Pr_i\right)$ for $1 \leq i \leq k$, and let $X:\Omega\rightarrow \mathbb{R}$ be a random variable with the property that the effect of the $i$-th coordinate is at most $c_i$. Then, 
\[
\Pr[X \le \mathbb{E}[X]-a]\leq\exp\left(-\frac{a^2}{2\sum_{i=1}^{k}c_{i}^2}\right).
\]
\end{theorem} 
Moreover, we sometimes use the basic inequalities $1-z\le \exp\left(-z\right)$ for any $z$ and $4^{-z}\le 1-z$ for any $0< z< 1/2$.

\paragraph{Assumptions.} All logarithms are to base $e$, otherwise we point out explicitly. Furthermore, we let $n$ (the number of nodes) tend to infinity and $\alpha, \epsilon\in (0,1/2)$ are constants. We assume that the error probability $\mu$ is larger than $1/\sqrt{n}$. (Actually, most of our proofs also work when $\mu>1/f(n)$ for any function $f(n)$ sub-exponential in $n$.)
\section{Strong Attacker}
\label{strong attacker}
In this section, our goal is to prove that a regular expander graph is resilient to a strong attacker. Roughly speaking, one says a graph has strong expansion properties if it is highly connected. There exist different parameters to measure the expansion of a graph. We consider an algebraic characterization of expansion. However, since the relation between other measures, such as vertex and edge expansion, and ours is well-understood, cf.~\cite{hoory2006expander}, our results can be immediately rephrased. 

Let $\sigma(G)$ be the
second-largest absolute eigenvalue of the normalized adjacency matrix of graph $G$. There is a rich literature about the relation between the value of $\sigma$ and the expansion of graph $G$. However for our purpose here, all one needs to know is that $G$ has stronger expansion properties when $\sigma$ is smaller. Now, we present the expander mixing lemma, cf.~\cite{hoory2006expander}, which basically states that the number of edges between any two node sets is almost completely determined by their cardinality if the value of $\sigma$ is small.
\begin{lemma}[Expander mixing lemma]
\label{mixing}
\label{general:mixinglemma} For a $d$-regular graph $G=\left(V,E\right)$ and $S,S'\subset V$
\[
\left|e\left(S,S'\right)-\frac{|S|\,|S'|d}{n}\right|\le\sigma d \,\sqrt{|S|\,|S'|}\,.
\]
\end{lemma}
\begin{theorem}
\label{strong-expander}
For a $d$-regular graph $G=(V,E)$ and an ($\alpha$,$\epsilon$)-strong attacker, if $\sigma\le \epsilon\sqrt{\alpha (1-\alpha)}$, then the attacker cannot win for any number of rounds.
\end{theorem}
\proof We know that $w_0=(1/2+\epsilon)\alpha n>\alpha n/2$. We prove that $w_1\ge(1-\alpha)n/2$, which yields $w_0+w_1>n/2$. Thus, for any number of rounds, white is the dominant color.

Let $A:=V\setminus (R_0\cup W_1)$ be the set of non-seed nodes which are black or uncolored after one round. It suffices to prove that $a:=|A|\le (1-\alpha)n/2$ (since this implies that $w_1\ge(1-\alpha)n/2$). For each node $v\in A$, $\deg_{W_0}(v)\le \deg_{B_0}(v)$ (because otherwise it would become white after one round). Hence, $e(A,W_0)\le e(A,B_0)$. By applying Lemma~\ref{mixing} to both sides of this inequality, we get
\[
\frac{aw_0d}{n}-\sigma d\sqrt{aw_0}\le \frac{ab_0d}{n}+\sigma d\sqrt{ab_0}.
\]
Dividing by $\sqrt{a}$ and re-arranging the terms give us
\[
\sqrt{a}(w_0-b_0)\le \sigma n(\sqrt{w_0}+\sqrt{b_0}).
\]
Since $\sqrt{w_0}+\sqrt{b_0}=\sqrt{(1/2+\epsilon)\alpha n}+\sqrt{(1/2-\epsilon)\alpha n}\le \sqrt{2\alpha n}$, $w_0-b_0=2\alpha\epsilon n$, and $\sigma\le \epsilon\sqrt{\alpha (1-\alpha)}$, we get
\[
\sqrt{a}\le \frac{\epsilon n \sqrt{\alpha (1-\alpha)} \sqrt{2\alpha n}}{2\alpha \epsilon n}\Rightarrow a\le (1-\alpha)n/2.\ \ \ \ \ \Box
\]
\paragraph{Random regular graphs.} The random $d$-regular graph $\mathcal{G}_{n,d}$ is the random graph with a uniform distribution over all $d$-regular graphs on $n$ nodes. It is proven by~\cite{friedman2003proof} that $\sigma(\mathcal{G}_{n,d})\le2/\sqrt{d}$ for $d\ge 3$ a.a.s. Putting this statement in parallel with Theorem~\ref{strong-expander} implies that if $d\ge 4/(\alpha (1-\alpha)\epsilon^2)$, then $\mathcal{G}_{n,d}$ is resilient to an ($\alpha$,$\epsilon$)-attacker a.a.s.

\paragraph{Irregular graphs.} So far, we limited ourselves to regular graphs. However, our result can be generalized to capture irregular graphs by applying basically the same proof ideas. All we need to do is to apply a more general variant of Lemma~\ref{mixing}, cf.~\cite{hoory2006expander}, and replace $d$ with $\delta$ or $\Delta$, according to the case. Then, we can conclude that for a graph $G=(V,E)$ and an ($\alpha$,$\epsilon$)-strong attacker, if $\sigma\le ((1+\gamma)\epsilon/2-(1-\gamma)/4)\sqrt{\alpha (1-\alpha)}$, then the attacker cannot win, where $\gamma:=\delta/\Delta$. (This is equivalent to the statement of Theorem~\ref{strong-expander} for $\gamma=1$.)

\paragraph{Erd\H{o}s-R\'{e}nyi random graph.} In the Erd\H{o}s-R\'{e}nyi random graph $\mathcal{G}_{n,p}$ each edge is added independently w.p. $p$ on a node set of size $n$. Let $K$ be a sufficiently large constant. It is proven by~\cite{le2017concentration} that if $p\ge K\log n/n$, then $\sigma(\mathcal{G}_{n,p})=\mathcal{O}(1/\sqrt{np})$ a.a.s. (recall that $\log n/n$ is the connectivity threshold). Furthermore, it is well known, cf.~\cite{dubhashi2009concentration}, that $\gamma\ge (1-\epsilon)$ for any $\epsilon>0$ a.a.s. if $p\ge K\log n/n$. Combining the last two statements and our proposition about irregular graphs implies that $\mathcal{G}_{n,p}$ for $p\ge K\log n/n$ is a.a.s. resilient to an ($\alpha$,$\epsilon$)-strong attacker.   

\paragraph{Tightness.} We believe that regularity and expansion are not only sufficient conditions for a graph to be resilient to a strong attacker, but also somehow necessary. To support this, in the appendix, Section~\ref{appendix-tightness-strong}, we provide a regular non-expander graph and an irregular expander graph where a strong attacker has a winning strategy.

\section{Moderate Attacker}
\label{moderate attacker}
In this section, our main purpose is to prove Theorem~\ref{moderate-attacker-Delta}. To do so, let us first provide Lemma~\ref{w0-size}.
\begin{lemma}
	\label{w0-size}
	For a graph $G=(V,E)$ and an ($\alpha$,$\epsilon$)-moderate attacker, $(1/2+\epsilon)\alpha n-\alpha\epsilon n/2\le w_0\le (1/2+\epsilon)\alpha n+\alpha\epsilon n/2$ w.p. $1-\exp(-\Theta(\epsilon^2\alpha n))$.
\end{lemma}
\proof Label the seed nodes arbitrarily from $v_1$ to $v_{\alpha n}$. We define Bernoulli random variable $x_i$ to be 1 if and only if $v_i$ is colored white. Then, $w_0=\sum_{i=1}^{\alpha n}x_i$ and $\mathbb{E}[w_0]=(1/2+\epsilon)\alpha n$. Since $x_i$s are independent, applying the Chernoff bound (Theorem~\ref{Chernoff}) for $\eta=\epsilon/2$ yields our claim.\qed
\begin{theorem}
\label{moderate-attacker-Delta}
For a graph $G=(V,E)$ and an ($\alpha$,$\epsilon$)-moderate attacker, if $\Delta\le (Cn/\log (1/\mu))^{\frac{1}{2t}}$ for some sufficiently small constant $C_{\alpha, \epsilon}>0$, then the attacker cannot win in $t$ rounds w.p. at least $1-\mu$.
\end{theorem}
\proof Let $r=\hat{r}_t-r_0$ be the number of nodes which are colored during rounds 1 to $t$. We define $X:=\hat{w}_t-w_0$ to be the number of nodes which are colored white among these nodes. We prove that $X> r/2-\alpha\epsilon n/2$ w.p. at least $1-\mu/2$. Furthermore based on Lemma~\ref{w0-size}, $w_0\ge r_0/2+\alpha\epsilon n/2$ w.p. $1-\exp(-\Theta(\epsilon^2\alpha n))\ge 1-1/(2\sqrt{n})\ge 1-\mu/2$ (where we used our assumptions that $\alpha, \epsilon$ are constant and $\mu\ge 1/\sqrt{n}$). Therefore, we have $\hat{w}_t=X+w_0> (r/2-\alpha\epsilon n/2)+(r_0/2+\alpha\epsilon n/2)=(r+r_0)/2=\hat{r}_t/2$ w.p. at least $1-\mu$, which implies that the attacker does not win in $t$ rounds.

It remains to prove that $X> r/2-\alpha\epsilon n/2$ w.p. at least $1-\mu/2$. Let us label all nodes in $R_0$ arbitrarily from $v_1$ to $v_{\alpha n}$. Corresponding to each node $v_i$ for $1\le i\le \alpha n$, we define the probability space $(\Omega_i,\Pr_i)$. Then, $X:\Omega\rightarrow \mathbb{R}$ is defined over $\Omega$, where $(\Omega,\Pr)$ is the product of discrete probability spaces $(\Omega_{i}, \Pr_{i})$. Recall that we say the effect of the $i$-th coordinate is at most $c_i$ if for all $\omega,\omega'\in \Omega$ which differ only in the $i$-th coordinate we have $|X(\omega)-X(\omega')|\le c_i$. Obviously, if we change the color of node $v_i$, then the color of at most $|\hat{N}_t(v_i)|\le 2\Delta^t$ of the nodes, which are colored in the first $t$ rounds, will be affected. Therefore, $c_i\le 2\Delta^t$, which implies that $\sum_{i=1}^{\alpha n}c_i^2\le 4\alpha n\Delta^{2t}$. Furthermore, $\mathbb{E}[X]\ge r/2$ since each node in $V\setminus R_0$ is white w.p. at least $1/2$ (actually, one can get the stronger lower bound of $(1/2+\epsilon)$ with a coupling argument, which is not needed here). Now, applying Azuma's inequality (Theorem~\ref{Azuma}) yields
\[
\Pr[X\le \frac{r}{2}-\frac{\alpha\epsilon n}{2}]\le \exp\left(-\frac{\alpha^2\epsilon^2n^2}{8\sum_{i=1}^{\alpha n}c_i^2}\right)\le \exp\left(-\frac{\alpha\epsilon^2 n}{32\Delta^{2t}}\right).
\]
For a suitable choice of $C_{\alpha,\epsilon}$, we have $\Delta\le \left(\frac{\alpha\epsilon^2 n}{32\log (2/\mu)}\right)^{\frac{1}{2t}}$. Thus, the above probability is at most $\mu/2$. \qed

\paragraph{Tightness.} Now, we provide Proposition~\ref{moderate-tightness} which asserts that if we could replace the exponent $1/2t$ with $1/t$ in Theorem~\ref{moderate-attacker-Delta}, then our bound would be tight, up to some constant factor. We believe that this is actually doable with some case distinctions and more careful calculations; however, we only prove such statement for $t=1$ here, in Proposition~\ref{moderate-Delta=t=1}. The proof of Propositions~\ref{moderate-tightness} and ~\ref{moderate-Delta=t=1} are given in the appendix, Sections~\ref{appendix-moderate-tightness} and~\ref{appendix-moderate-Delta=t=1}.   

\begin{proposition}
\label{moderate-tightness}
For any $\epsilon, \alpha, \mu>0$ and $t\in \mathbb{N}$, there exists a graph $G$ with $\Delta=(C^{\prime}n/\log(1/\mu))^{\frac{1}{t}}$ for some constant $C^{\prime}_{\alpha, \epsilon}$ such that an ($\alpha$,$\epsilon$)-moderate attacker can win in $t$ rounds w.p. larger than $\mu$.
\end{proposition}

\begin{proposition}
\label{moderate-Delta=t=1}
For a graph $G=(V,E)$ and an ($\alpha$,$\epsilon$)-moderate attacker, if $\Delta\le C^{\prime\prime}n/\log^2 (1/\mu)$ for some sufficiently small constant $C_{\alpha, \epsilon}^{\prime\prime}>0$, then the attacker cannot win in one round w.p. at least $1-\mu$.
\end{proposition}

\section{Weak Attacker}
\label{weak attacker}
Theorem~\ref{Delta-weak} states that if the maximum degree of a graph is sufficiently smaller than $n$ or if the degree of at least half of the nodes is pretty large, then it is resilient to a weak attacker. In other words, a weak attacker can be successful only on graphs with some nodes of very high degree and a large set of nodes with small degree such as a star graph.
\begin{theorem}
\label{Delta-weak}
For a graph $G=(V,E)$ and an ($\alpha$,$\epsilon$)- weak attacker, if $\Delta\le Cn/((\log n)^{C^{\prime}}\log (4/\mu))$ for some suitable constants $C_{\alpha, \epsilon}, C_{\alpha,\epsilon}^{\prime}>0$ or if half of the nodes are of degree at least $d^*_1:=(6/\alpha\epsilon^2)\log (12/\alpha\epsilon \mu)$, then for any $t\in\mathbb{N}$ the attacker cannot win in $t$ rounds w.p. at least $1-\mu$.
\end{theorem}
We observe that the bound on $\Delta$ in Theorem~\ref{Delta-weak} does not depend on the number of rounds $t$, unlike the case of a moderate attacker. Furthermore for a moderate attacker, we cannot prove that if the degree of half of the nodes is larger than a certain degree threshold, then the attacker fails, except if the threshold is very large. Assume that graph $G$ is the union of a clique of size $\alpha n-1$ and a clique of size $(1-\alpha)n+1$. Suppose that an ($\alpha$,$\epsilon$)-moderate attacker chooses all $\alpha n-1$ nodes in the first clique and a node $v$ in the second one to be the seed nodes. If $v$ is colored black, then the attacker wins and this happens w.p. $(1/2-\epsilon)$. Note that all nodes in $G$ are of degree at least $\alpha n-2$. 

\paragraph{Proof sketch.} The complete proof is fairly long and is given in the appendix, Section~\ref{appendix-Delta-weak}. Here, we deliver some key ideas which the proof is built on. 

Since each of a node's neighbors is initially black independently w.p. $(1/2-\epsilon)\alpha$ and white w.p. $(1/2+\epsilon)\alpha$, the dominant color in the neighborhood of a high-degree node is very likely to be white. This lets us prove that if half of the nodes are of degree at least $d^*_1$, then after one round more than $n/2$ nodes are white w.p. at least $1-\mu$. In that case, the attacker cannot win after any number of rounds.

For the case of $\Delta\le Cn/((\log n)^{C^{\prime}}\log (4/\mu))$, we first show that $w_0\ge b_0+\alpha\epsilon n$ w.p. at least $1-\mu/4$, similar to the proof of Lemma~\ref{w0-size}. Furthermore, one can prove that at most $\alpha\epsilon n/2$ nodes are colored after the $t^*_1$-th round w.p. at least $1-\mu/4$, for $t^*_1:=(2/\alpha)\log (4/\alpha\epsilon)$. In the worst case scenario, all these nodes are colored black, but this is not an issue since we already have $\alpha\epsilon n$ extra white nodes in $R_0$. Moreover building on the argument from the previous paragraph, we can prove that all non-seed nodes whose degree is larger than $d^*_2:=8\log n/(\alpha\epsilon^2)$ are colored white after one round w.p. at least $1-\mu/4$. Therefore, it only remains to show that at least half of the nodes whose degree is smaller than $d^*_2$ and are in $\cup_{i=1}^{t^*_1} R_i$ will be colored white w.p. at least $1-\mu/4$. For this, we rely on Azuma's inequality, similar to the proof of Theorem~\ref{moderate-attacker-Delta}. However, we can prove a much tighter bound on $\sum_{i=1}^{r_0}c_i^2$ by a smarter counting argument and using the fact that we only need to focus on ``low-degree'' nodes that are colored up to the $t^*_1$-th round.~\qed

\paragraph{Tightness.} The bound $(6/\alpha\epsilon^2)\log (12/\alpha\epsilon \mu)$ in Theorem~\ref{Delta-weak} is tight in terms of $\mu$. However, we believe the dependency on $\alpha$ and $\epsilon$ is not best possible. The upper bound on $\Delta$ is also tight in $\mu$ and the dependency on $n$ is optimal, up to the poly-logarithmic term. Please see Section~\ref{appendix-weak-tightness} for more details including the proof of these claims.
\section{Stabilization Time}
\label{stabilization time}
In this section, we prove tight bounds on the stabilization time of our opinion diffusion process. Let $G$ be a connected graph. (Otherwise, we just need to consider the maximum stabilization time among all the connected components.) Note that stabilization time is only a function of the choice of seed nodes, not their color. After $t$ rounds, all nodes whose distance from the seed set is at most $t$ will be colored. Therefore, the stabilization time is upper-bounded by the diameter $D(G)$, which is the greatest distance between any pair of nodes in $G$. This bound is obviously tight. Consider two nodes $v,u$ such that $d(v,u)=D$. If $v$ is the only seed node, then it takes $D$ rounds until node $u$ is colored.

What if the seed set is chosen randomly (as in case of a weak attacker)? We prove that if each node is chosen to be a seed node independently w.p. $\alpha$, then a.a.s. the stabilization time is between $\Omega(\log_{\Delta}\log n^{\frac{1}{\alpha}})$ (see Theorem~\ref{stabilization-time-lower}) and $\mathcal{O}((1/\alpha\delta)\log n)$ (see Theorem~\ref{stabilization-time-random}).

\begin{theorem}
	\label{stabilization-time-random}
	For a graph $G=(V,E)$, if each node is a seed node independently w.p. $\alpha$, then the stabilization time is in $\mathcal{O}((1/\alpha\delta)\log n)$ a.a.s. 
\end{theorem}
To prove Theorem~\ref{stabilization-time-random}, we first provide Lemmas~\ref{neighborhood-lower-bound} and~\ref{logn-coin_flip}.

\begin{lemma}
	\label{neighborhood-lower-bound}
	For a node $v$ in a graph $G=(V,E)$ and $t\in \mathbb{N}$, if $N_t(v)\ne \emptyset$, then $|\hat{N}_t(v)|\ge (t-1)\delta/3$.
\end{lemma}
\proof Consider a triple $(N_{3i}(v),N_{3i+1}(v), N_{3i+2}(v))$ for some $0\le i\le \lfloor (t+1)/3\rfloor-1$. Since $N_{3i+1}(v)$ is non-empty, it must include a node $u$. All the neighbors of $u$ are in this triple, i.e., $N(u)\subset N_{3i}(v)\cup N_{3i+1}(v)\cup N_{3i+2}(v)$. Since $\deg (u)=|N(v)|\ge \delta$, the triple must include at least $\delta+1$ nodes. Thus, we get $|\hat{N}_t(v)|\ge \lfloor \frac{t+1}{3}\rfloor (\delta+1)\ge(t-1)\delta/3$.~\qed

\begin{lemma}
	\label{logn-coin_flip}
	Assume each node in a graph $G=(V,E)$ is a seed node independently w.p. $\alpha$. Then, a.a.s. there is no node $v$ such that $|\hat{N}_t(v)|\ge (2/\alpha)\log n$ and $\hat{N}_t(v)\cap R_0=\emptyset$.
\end{lemma}~\proof The probability that $\hat{N}_t(v)\cap R_0=\emptyset$ is equal to 
\[
(1-\alpha)^{\left|\hat{N}_t(v)\right|}\le (1-\alpha)^{\frac{2\log n}{\alpha}}\le \exp(-2\log n)=\frac{1}{n^2}.
\]
Thus, a union bound implies that w.p. $1-1/n$ there is no node $v$ such that $|\hat{N}_t(v)|\ge \frac{2}{\alpha}\log n$ and $\hat{N}_t(v)\cap R_0=\emptyset$.~\qed 

\paragraph{Proof of Theorem~\ref{stabilization-time-random}.} Let $t^*_2:=(6/\alpha\delta)\log n+1$. If $N_{t^*_2}(v)=\emptyset$ for a node $v\in V$, it will be colored in less than $t^*_2$ rounds or it will never be colored. On the other hand, if $N_{t^*_2}(v)\ne\emptyset$, by Lemma~\ref{neighborhood-lower-bound} we have $|\hat{N}_{t^*_2}(v)|\ge (t^*_2-1)\delta/3=(2/\alpha)\log n$. By Lemma~\ref{logn-coin_flip}, we know that a.a.s. every node whose $t^*_2$-neighborhood is of size at least $(2/\alpha)\log n$ has a seed node in its $t^*_2$-neighborhood, and thus is colored in at most $t^*_2$ rounds. Hence, a.a.s. after $t^*_2=(6/\sigma\alpha)\log n+1=\mathcal{O}((1/\alpha\delta)\log n)$ rounds every node is colored or it will never be colored. \qed 

\paragraph{Tightness.} Let $C_n^{\delta}$ be the $\delta$-th power of a cycle $C_n$, which is a $2\delta$-regular graph. We can prove that if each node is a seed node independently w.p. $\alpha$, then a.a.s. there is a node $u$ such that $\hat{N}_{t^*_3}(u)\cap R_0=\emptyset$, for $t^*_3:=(1/16\alpha\delta)\log_2 n-1$, which implies that it will be colored after the $t_3^*$-th round. Thus, the process takes at least $t^*_3=\Omega((1/\alpha\delta)\log n)$ rounds to end. A complete proof is given in the appendix, Section~\ref{appendix-thigtness-1}.
\begin{theorem}
	\label{stabilization-time-lower}
	For a connected graph $G=(V,E)$, if each node is a seed node independently w.p. $\alpha$, then the stabilization time is in $\Omega(\log_{\Delta}\log n^{\frac{1}{\alpha}})$ a.a.s.
\end{theorem}
\paragraph{Tightness.} We can prove that if each node in a complete $\Delta$-ary tree is a seed node independently w.p. $\alpha$, then a.a.s. after $\mathcal{O}(\log_{\Delta}\log n^{\frac{1}{\alpha}})$ rounds all nodes are colored. This implies that the lower bound in Theorem~\ref{stabilization-time-lower} is tight. The proof of this statement and Theorem~\ref{stabilization-time-lower} are given in the appendix, Sections~\ref{appendix-thigtness-2} and~\ref{appendix-stabilization-time-lower} respectively.
\section{Hardness Results}
\label{hardness results}
Assume you are given a graph $G=(V,E)$ and some $t\in \mathbb{N}$. The goal is to determine whether there is a choice of the seed set for which the process needs exactly $t$ rounds to end. This problem is polynomial-time solvable. If $t>D$, where $D$ is the diameter of $G$, then the answer is No since the stabilization time is upper-bounded by $D$. If $t\le D$, then the answer is Yes. Assume that $d(v,u)=D$ for some nodes $v,u\in V$. Then, if the seed set is equal to $\hat{N}_{D-t}(v)$, the process takes exactly $t$ rounds. However, if in the above problem we require that the seed set must be of a given size $s$, then the problem becomes NP-hard. It is because the \textit{$t$-hop dominating set problem}, which is proven to be NP-hard by~\cite{basuchowdhuri2014finding}, can be reduced to this problem. Assume we are given a graph $G$ and integers $t, s$ as input. In the $t$-hop dominating set problem, the task is to decide whether there is a node set $S\subseteq V$ of size $s$ such that $d(S,v)\le t$ for any $v\in V$. We observe that the answer to this problem is Yes if and only if there is a seed set of size $s$ for which the process takes at most $t$ rounds. Hence, this problem can be reduced to our stabilization time problem.  

Now, assume that you are given an $n$-node graph $G$ and some integers $b$, $w$, and $t$. Your task is to choose a seed set of $b$ black nodes and $w$ white nodes such that $\hat{w}_t$ is as small as possible. (This is similar to what a strong attacker aims to do.) We call this the Minimum Influence (MI) problem.
\begin{theorem}
\label{hardness-MI-problem}
Let $\mathcal{A}$ be a polynomial-time $\beta$-approximation algorithm for the MI problem. Then, $\beta>n^{1-\zeta}$ for any $\zeta>0$ unless P=NP.
\end{theorem}
\paragraph{Proof sketch.} In the \textit{clique problem}, which is NP-hard, a graph $G'$ and an integer $k$ are given and the task is to determine whether $G'$ has a clique of size $k$. Let $G'$ and $k$ be the input of the clique problem. Then, we construct an $n$-node graph $G$ such that the solution of the MI problem for $b=k$, $w={k\choose 2}$ on $G$ is ${k\choose 2}$ if $G'$ has a clique of size $k$ and it is larger than $n^{1-\zeta} {k\choose 2}$ if $G'$ does not. This yields our claim.

To construct $G$, we essentially replace each edge and node in a copy of $G'$ with some gadget, which includes a large clique. Our construction is tailored in a way that if $G'$ has a clique of size $k$, then we can place $w={k\choose 2}$ white nodes on the gadgets corresponding to the edges of this clique such that no new white node will be generated. However, when there is no clique of size $k$, then a large group of nodes will become white after two rounds, no matter how we place the $w$ white nodes. A complete proof is given in Section~\ref{appendix-hardness-MI-problem}. \qed

\section{Conclusion}
\label{conclusion}
It would be interesting to study other variants of our model or attackers. Our results apply to some other settings with some minor changes. For example, our propositions regarding the attackers should still hold if in each round only a random subset of nodes update their color or if each node does not follow the majority rule with some noise probability smaller than $\epsilon$. However, our proof techniques cannot immediately be applied to characterize graph structures resilient to an attacker which can choose the order of updates or add/delete edges. Hence, this is a potential avenue for further research.

The maximum degree in most of real-world social graphs is not too large, which makes them fairly resilient to a moderate and weak attacker. However, they are usually far from being regular or expander, and thus are vulnerable to a strong attacker. A counter-measure might be to add an Erd\H{o}s-R\'{e}nyi random graph on top of the underlying graph structure. In the real world, this is essentially the same as asking agents to choose a fraction of their connections at random. Our preliminary experiments on data from SNAP~\cite{snapnets} suggest that such modification enhances the resilience significantly. However, a detailed and formal investigation of this subject is left for future research.

At first glance, our inapproximability result in Theorem~\ref{hardness-MI-problem} might create the confusion that a strong attacker is computationally impotent. However, it is possible that there is an optimal strategy for special classes of graphs; in particular, for the ones which commonly occur in the real world. Furthermore, assume that in the MI problem, the seed set is given as a part of the input and the attacker's task is only to decide on the color of the seed nodes. Then, the problem can be reduced to the \textit{partial set multi-cover problem}~\cite{shi2019approximation}, which has a constant factor approximation algorithm. Therefore, if the attacker has an effective heuristic method to select the seed set, then it can find a reasonably efficient solution. Further investigation of the computational power of an attacker in our model is a prospective research direction.
\section{ Acknowledgments}
The author is thankful to V\'{a}clav Rozho\v{n} and Christoph Grunau for several stimulating conversations, especially on Theorem~\ref{hardness-MI-problem}.
\bibliographystyle{acm}
\bibliography{ref}
\appendix
\section{Tightness of Theorem~\ref{strong-expander}}
\label{appendix-tightness-strong}
As mentioned, we believe that regularity and expansion are not only sufficient conditions for a graph to be resilient to a strong attacker, but also somehow necessary. To support this, we give a regular non-expander and an irregular expander graph where a strong attacker has a winning strategy.

Let $\alpha$ and $\epsilon$ be arbitrary constants between $0$ and $1/2$ and assume that $d\ge 2$. Consider an arbitrary $d$-regular graph $G_1$ on $(1/2+\epsilon)\alpha n$ nodes and a $d$-regular connected graph $G_2$ on $n-(1/2+\epsilon)\alpha n$ nodes. Let $G$ be the union of $G_1$ and $G_2$. Thus, $G$ is an $n$-node $d$-regular graph. An ($\alpha$, $\epsilon$)-strong attacker can simply color all nodes in $G_1$ white and $(1/2-\epsilon)\alpha n$ arbitrary nodes in $G_2$ black. Then, the number of white nodes remains unchanged while black color eventually takes over all uncolored nodes. Therefore, the attacker wins.

Now, we provide an irregular expander graph which is not resilient to a strong attacker. Let $\alpha,\epsilon>0$ be sufficiently small constants and assume that $G$ is a $d$-regular graph, for some fixed $d$, with strong expansion properties. Let $B$ and $W$ be two arbitrary disjoint node sets of size $(1/2-\epsilon)\alpha n$ and $(1/2+\epsilon)\alpha n$, respectively. Now, we add $d$ additional edges between each node $v\in B$ and nodes in $V\setminus (B\cup W)$ at random. One can show that this new graph has strong expansion properties. However, if an ($\alpha$, $\epsilon$)-attacker chooses $B_0=B$ and $W_0=W$, then a.a.s. each black node ``generates'' almost $2d$ new black nodes in the next round while each white node generates almost $d$ new white nodes a.a.s., i.e., $b_1\simeq 2db_0$ and $w_1\simeq dw_0$. Thus, the number of black nodes surpasses the number of white ones after one round. (We should stress that the given argument is not a formal proof since we left out several details. However, our goal here is only to provide an intuitive argument.) 

\section{Proof of Proposition~\ref{moderate-tightness}}
\label{appendix-moderate-tightness}
To prove Proposition~\ref{moderate-tightness}, we use Markov's inequality, cf.~\cite{dubhashi2009concentration}.
\begin{theorem}[Markov's inequality]
\label{Markov}
Let $X$ be a non-negative random variable with finite expectation and $a>0$, then
\[
\Pr[X\ge a]\le \frac{\mathbb{E}[X]}{a}.
\]
\end{theorem}
\begin{customp}{\ref{moderate-tightness}}
For any $\epsilon, \alpha, \mu>0$ and $t\in \mathbb{N}$, there exists a graph $G$ with $\Delta=(C^{\prime}n/\log(1/\mu))^{\frac{1}{t}}$ for some constant $C^{\prime}_{\alpha, \epsilon}$ such that an ($\alpha$,$\epsilon$)-moderate attacker can win in $t$ rounds w.p. larger than $\mu$.
\end{customp}
\proof 
The proof is essentially built on the simple fact that if in a complete $d$-ary tree the root $v$ is black (and the rest of nodes are uncolored), then after $t$ rounds $|\hat{N}_t(v)|\ge d^t$ nodes will be black.

Let $s=\log_2(1/\mu)/(2+8\epsilon)$. Assume graph $G$ is the union of a cycle on $\alpha n-s$ nodes and $s$ complete $d$-ary trees, each of size $((1-\alpha)n+s)/s$. Suppose that $$d=((3\alpha\epsilon n+1)/s)^{\frac{1}{t}}=((2+8\epsilon)(3\alpha\epsilon n+1)/\log_2(1/\mu))^{\frac{1}{t}}.$$ Thus $G$ is an $n$-node graph with maximum degree $d+1$, which is equal to $\Delta=(C^{\prime}_{\alpha, \epsilon}n/\log(1/\mu))^{\frac{1}{t}}$ for a suitable choice of $C_{\alpha,\epsilon}^{\prime}$.

Assume that the attacker chooses the $\alpha n-s$ nodes on the cycle and $s$ roots of the $d$-ary trees to be the seed nodes. Then, the probability that all $s$ roots are colored black is
\[
\left(\frac{1}{2}-\epsilon\right)^s=2^{-s}\left(1-2\epsilon\right)^s\ge 2^{-s}4^{-2\epsilon s}=\sqrt{\mu}>2\mu
\] 
where we used $(1-z)\ge 4^{-z}$ for $0<z<1/2$ and $s=\log_2(1/\mu)/(2+8\epsilon)$, respectively.

Furthermore, if all $s$ roots are black, then at least $sd^t=3\alpha\epsilon n+1$ uncolored nodes will be colored black after $t$ rounds (while no uncolored node will be colored white). Hence, w.p. more than $2\mu$, the number of black nodes increases by at least $3\alpha\epsilon n+1$ while the number of white ones remains unchanged. Moreover based on Lemma~\ref{w0-size}, $w_0\le b_0+3\alpha\epsilon n$ w.p. $1-\exp(-\Theta(\epsilon^2\alpha n))>1-1/\sqrt{n}\ge 1-\mu$, where we sued our assumptions that $\alpha,\epsilon$ are constant and $\mu\ge 1/\sqrt{n}$. Therefore, we can conclude that w.p. more than $1-(1-2\mu)-\mu=\mu$, we have $\hat{b}_t>\hat{w}_t$, that is, the attacker wins.\qed

\section{Proof of Proposition~\ref{moderate-Delta=t=1}}
\label{appendix-moderate-Delta=t=1}
\begin{customp}{\ref{moderate-Delta=t=1}}
For a graph $G=(V,E)$ and an ($\alpha$,$\epsilon$)-moderate attacker, if $\Delta\le C^{\prime\prime}n/\log^2 (1/\mu)$ for some sufficiently small constant $C_{\alpha, \epsilon}^{\prime\prime}>0$, then the attacker cannot win in one round w.p. at least $1-\mu$.
\end{customp}
\proof Suppose that $d^*_3:=4\log (6/\mu)/\epsilon^2$. We define $$R_{1h}:=\{v\in R_1: \deg_{R_0}(v)\ge d^*_3\}$$ to be the set of \emph{high-degree} nodes in $R_1$ and $R_{1l}:=R_1\setminus R_{1h}$ to be the set of \emph{low-degree} nodes in $R_1$. Let $r_{1h}:=|R_{1h}|$ and $r_{1l}:=|R_{1l}|$. We also let $w_{1h}$ and $w_{1l}$ denote the number of white nodes in $R_{1h}$ and $R_{1l}$, respectively. Analogously, we define, $b_{1h}$ and $b_{1l}$ for black color.

First, we show that $w_{1h}\ge r_{1h}/2$ w.p. at least $1-\mu/3$. Then, we prove that $w_{1l}> r_{1l}/2-\alpha\epsilon n/2$ w.p. at least $1-\mu/3$. Furthermore based on Lemma~\ref{w0-size}, we know that $w_0\ge r_0/2+\alpha\epsilon n/2$ w.p. $1-\exp(-\Theta(\epsilon^2\alpha n))\ge 1-1/3\sqrt{n}\ge 1-\mu/3$ (where we used our assumptions that $\alpha,\epsilon$ are constant and $\mu\ge 1/\sqrt{n}$). Combining these three statements implies that $\hat{w}_1=w_0+w_{1h}+w_{1l}> (r_0+r_{1h}+r_{1l})/2=\hat{r}_1/2$ (i.e., the attacker does not win after one round) w.p. at least $1-\mu$. 

Let us first prove that $w_{1h}\ge r_{1h}/2$ w.p. at least $1-\mu/3$. Consider an arbitrary node $v$ in $R_{1h}$ and let $u_i$ for $1\le i\le d=\deg_{R_0}(v)$ be the neighbors of $v$ in $R_0$. We define Bernoulli random variable $x_i$ to be 1 if and only if $u_i$ is colored white. Let $X:=\sum_{i=1}^{d}x_i$ denote the number of white neighbors of $v$ in $R_0$. Thus, we have $\mathbb{E}[X]=(1/2+\epsilon)d$. Then, the probability that $v$ is colored black is at most
\begin{align*}
\Pr[X\le d/2]&\le \Pr[X\le (1-\epsilon)(1/2+\epsilon)d] \\ &= \Pr[X\le (1-\epsilon)\mathbb{E}[X]]\\ &\le \exp(-\epsilon^2(1/2+\epsilon)d/2)\\ &\le \exp(-\epsilon^2d^*_3/4)\\ &=\exp(-\log (6/\mu))\\ &=\mu/6
\end{align*}
where we used the Chernoff bound (Theorem~\ref{Chernoff}). Therefore, the probability that an arbitrary node in $R_{1h}$ is colored black is at most $\mu/6$. This implies that $\mathbb{E}[b_{1h}]\le r_{1h}\ \mu/6$. Applying Markov's inequality (Theorem~\ref{Markov}) gives us
\begin{align*}
\Pr[w_{1h}< \frac{r_{1h}}{2}]&=\Pr[b_{1h}\ge \frac{r_{1h}}{2}]\\ &\le \frac{\mathbb{E}[b_{1h}]}{r_{1h}/2}\\ &\le \frac{r_{1h}\ \mu/6}{r_{1h}/2}\\ & =\mu/3.
\end{align*}
It remains to prove that $w_{1l}> r_{1l}/2-\alpha\epsilon n/2$ w.p. at least $1-\mu/3$. 
Let $v_i$ for $1\le i\le \alpha n$ be the nodes in $R_0$. Corresponding to each node $v_i$, we define the probability space $(\Omega_i,\Pr_i)$. Then, random variable $w_{1l}:\Omega\rightarrow \mathbb{R}$ is defined over $\Omega$, where $(\Omega,\Pr)$ is the product of discrete probability spaces $(\Omega_{i}, \Pr_{i})$ for $1\le i\le \alpha n$. Recall that we say the effect of the $i$-th coordinate is at most $c_i$ if for all $\omega,\omega'\in \Omega$ which differ only in the $i$-th coordinate we have $|w_{1l}(\omega)-w_{1l}(\omega')|\le c_i$. We claim that $c_i\le \deg_{R_{1l}}(v_i)$ because if we change the color of node $v_i$, then the only nodes in $R_{1l}$ which might adopt a different color are the nodes in $N_{R_{1l}}(v_i)$. Therefore, we have that
\[
\sum_{i=1}^{\alpha n}c_i \le \sum_{i=1}^{\alpha n}\deg_{R_{1l}}(v_i)=\sum_{u\in R_{1l}}\deg_{R_0}(u)\le r_{1l}d^*_3.
\]
Using $d^*_3=4\log (6/\mu)/\epsilon^2$ and $r_{1l}\le n$ we get
\[
\sum_{i=1}^{\alpha n}c_i\le \frac{4n\log \frac{6}{\mu}}{\epsilon^2}.
\]
Since $c_i\le \deg_{R_{1l}}(v_i)\le \Delta$, we have
\[
\sum_{i=1}^{\alpha n}c_i^2\le \Delta \sum_{i=1}^{\alpha n}c_i\le \frac{4n\Delta\log \frac{6}{\mu}}{\epsilon^2}.
\]

Furthermore, $\mathbb{E}[w_{1l}]\ge r_{1l}/2$ since each node in $R_{1l}$ is white w.p. at least $1/2$. Now, applying Azuma's inequality (Theorem~\ref{Azuma}) yields
\begin{align*}
\Pr[w_{1l}\le \frac{r_{1l}}{2}-\frac{\alpha\epsilon n}{2}]&\le \exp\left(-\frac{\alpha^2\epsilon^2n^2}{8\sum_{i=1}^{\alpha n}c_i^2}\right)\\ &\le \exp\left(-\frac{\alpha^2\epsilon^4 n}{32\Delta\log\frac{6}{\mu}}\right) \\ & \le\exp\left(-\log \left(3/\mu\right)\right)\\ &=\mu/3.
\end{align*}
In the second step from last, we used that for a suitable choice of $C_{\alpha,\epsilon}^{\prime\prime}$ we have
\[
\Delta\le \frac{C^{\prime\prime}n}{\log^2 \frac{1}{\mu}}= \frac{\alpha^2\epsilon^4 n}{32\log \frac{6}{\mu}\log \frac{3}{\mu}}.
\]~\qed

\section{Proof of Theorem~\ref{Delta-weak}}
\label{appendix-Delta-weak}
In this section, our goal is to prove Theorem~\ref{Delta-weak}. We first present some necessary ingredients in Lemmas~\ref{w_0-size-weak},~\ref{high-degree-weak},~\ref{time-small-weak}, and~\ref{delta-half-weak}.
\begin{lemma}
	\label{w_0-size-weak}
	For a graph $G=(V,E)$ and an ($\alpha$,$\epsilon$)-weak attacker, $r_0/2+\alpha\epsilon n/2\le w_0 \le r_0/2+3\alpha\epsilon n/2$ w.p. at least $1-\exp(-\Theta(\epsilon^2\alpha n))$.
\end{lemma}
\proof We prove the lower bound. The proof of the upper bound is very similar. We know that $r_0/2+\alpha\epsilon n/2\le w_0$ is the same as $b_0+\alpha\epsilon n\le w_0$ because $r_0=w_0+b_0$. Thus, it suffices to prove that $w_0\ge (1/2+\epsilon/2) \alpha n$ and $b_0\le (1/2-\epsilon/2)\alpha n$ w.p. $1-\exp(-\Theta(\epsilon^2\alpha n))$. We prove the former and the latter is proven analogously. Assume that $v_1,\cdots, v_n$ are the nodes of $G$. We define Bernoulli random variable $x_i$ to be 1 if and only if node $v_i$ is chosen as a seed node and is colored white. Then, we have $\Pr[x_i=1]=(1/2+\epsilon)\alpha$ and $w_0=\sum_{i=1}^{n}x_i$. Thus, we get $\mathbb{E}[w_0]=(1/2+\epsilon)\alpha n$. Applying the Chernoff bound (Theorem~\ref{Chernoff}) yields
\begin{align*}
\Pr[w_0\le \left(\frac{1}{2}+\frac{\epsilon}{2}\right)\alpha n] &\le \Pr[w_0\le\left(1-\frac{\epsilon}{2}\right)\left(\frac{1}{2}+\epsilon\right)\alpha n]\\ &= \Pr[w_0\le \left(1-\frac{\epsilon}{2}\right)\mathbb{E}[w_0]]\\ & = \exp\left(-\Theta\left(\epsilon^2\alpha n\right)\right). 
\end{align*} \qed
\begin{lemma}
\label{high-degree-weak}
For a graph $G=(V,E)$ and an ($\alpha$,$\epsilon$)-weak attacker, every node whose degree is at least $d^*_2:=8\log n/(\alpha\epsilon^2)$ is in $R_0\cup W_1$ w.p. at least $1-2/n$.
\end{lemma}
\proof Consider an arbitrary node $v$ such that $\deg(v)\ge d^*_2$. Let $X_w:=|N(v)\cap W_0|$ and $X_b:=|N(v)\cap B_0|$. Then, $$\Pr[v\notin R_0\cup W_1]\le(1-\alpha)\Pr[X_w\le X_b].$$ We prove that $\Pr[X_w\le X_b]\le 2/n^2$. Thus, a union bound yields our desired statement.

Let $d=\deg(v)$. We prove that $X_w\le \alpha dn/2$ w.p. at most $1/n^2$. Analogously, one can show that $X_b\ge \alpha dn/2$ w.p. at most $1/n^2$. Thus, we have $X_w\le X_b$ w.p. at most $2/n^2$. Let $u_i$ for $1\le i\le d$ be the neighbors of $v$. We define Bernoulli random variable $x_i$ to be 1 if and only if $u_i$ is selected as a seed node and is colored white, which implies that $X_w=\sum_{i=1}^{d}x_i$. Since $\Pr[x_i=1]=(1/2+\epsilon)\alpha$, we get $\mathbb{E}[X_w]=(1/2+\epsilon)\alpha d$. Now, applying the Chernoff bound (Theorem~\ref{Chernoff}) yields
\begin{align*}
\Pr[X_w\le \alpha d/2]&\le \Pr[(1-\epsilon)(1/2+\epsilon)\alpha d] \\ &=\Pr[(1-\epsilon)\mathbb{E}[X_w]]\\ &\le \exp(-\epsilon^2 (1/2+\epsilon)\alpha d/2)\\ &\le \exp(-\epsilon^2\alpha d^*_2/4)\\ &=1/n^2
\end{align*}
where we used $d\ge d^*_2=8\log n/(\epsilon^2\alpha)$. \qed
\begin{lemma}
\label{time-small-weak}
For a graph $G=(V,E)$ and an ($\alpha$,$\epsilon$)-weak attacker, the number of nodes which are colored after the $t^*_1$-th round, for $t^*_1:=(2/\alpha)\log (4/\alpha\epsilon)$, is at most $\alpha\epsilon n/2$ w.p. at least $1-1/4\sqrt{n}$.
\end{lemma}
\proof If $N_{t^*_1}(v)=\emptyset$ for a node $v\in V$, then it will be colored in the first $t^*_1-1$ rounds or never will be colored. Thus, we only need to focus on nodes in $$V^{\prime}:=\{v\in V: N_{t^*_1}(v)\ne \emptyset\}.$$Let us define $$V_h^{\prime}:=\{v\in V^{\prime}:|\hat{N}_{t^*_1}(v)|\ge (2/\alpha)\log n\}$$ and $V_l^{\prime}:=V^{\prime}\setminus V_h^{\prime}$. Then, we let
\[
X_h:=|\{v\in V_h^{\prime}:\hat{N}_{t^*_1}(v)\cap R_0=\emptyset\}|
\]
and 
\[
X_l:=|\{v\in V_l^{\prime}:\hat{N}_{t^*_1/2}(v)\cap R_0=\emptyset\}|.
\]
We prove that $X_h=0$ w.p. at least $1-1/8\sqrt{n}$ and $X_l\le \alpha\epsilon n/2$ w.p. at least $1-1/8\sqrt{n}$. We observe that if there is a seed node in distance at most $t^*_1$ (subsequently, $t^*_1/2$) of a node, then it will be colored in at most $t^*_1$ rounds. Hence, we can conclude that there are at most $\alpha \epsilon n/2$ nodes which are colored after the $t^*_1$-th round w.p. at least $1-1/4\sqrt{n}$.

Consider an arbitrary node $v\in V_h^{\prime}$. Since $|\hat{N}_{t^*_1}(v)|\ge (2/\alpha)\log n$, then
\begin{align*}
\Pr[\hat{N}_{t^*_1}(v)\cap R_0=\emptyset]&\le (1-\alpha)^{(2/\alpha)\log n}\\ &\le \exp(-2\log n)\\ &=1/n^2.
\end{align*}
A union bound implies that $X_h=0$ w.p. at least $1-1/n\ge 1-1/8\sqrt{n}$.

Now, we prove that $X_l\le \alpha\epsilon n/2$ w.p. at least $1-1/8\sqrt{n}$. Let $u_1, \cdots, u_k$ be the nodes in $V_l^{\prime}$, where $k=|V_l^{\prime}|$. We define Bernoulli random variable $x_i$ to be 1 if and only if $\hat{N}_{t^*_1/2}(u_i)\cap R_0=\emptyset$. Thus, we have $X_l=\sum_{i=1}^{k}x_i$. Furthermore, $|\hat{N}_{t^*_1/2}(u_i)|\ge t^*_1/2$ because $\hat{N}_{t^*_1/2}(u_i)\ne \emptyset$. This implies that
\begin{align*}
\Pr[x_i=1]&\le (1-\alpha)^{t^*_1/2}\\ &\le \exp(-t^*_1\alpha/2)\\ &=\exp(-\log (4/\alpha\epsilon))\\ &=\alpha\epsilon/4.
\end{align*}
Therefore, we have
\begin{equation}
\label{equation 1-expectation}
\mathbb{E}[X_l]\le \alpha\epsilon k/4\le \alpha\epsilon n/4.
\end{equation}

Now, the goal is to prove our claim by using Chebyshev's inequality, cf.~\cite{dubhashi2009concentration}, which states that for a random variable $X$ with finite variance and $a>0$, we have $$\Pr[X\ge \mathbb{E}[X]+a]\le \frac{Var[X]}{a^2}.$$ Thus, we need to compute $\textrm{Var}[X_l]$. We have
\[
\textrm{Var}[X_l]=\sum_{i=1}^{k}\textrm{Var}[x_i]+\sum_{1\le i\ne j\le k} \textrm{Cov}(x_i,x_j).
\]
If $\hat{N}_{t^*_1/2}(u_i)\cap \hat{N}_{t^*_1/2}(u_j)=\emptyset$, then random variables $x_i$ and $x_j$ are independent, which implies that $\textrm{Cov}(x_i,x_j)=0$. On the other hand, if $\hat{N}_{t^*_1/2}(u_i)\cap \hat{N}_{t^*_1/2}(u_j)\ne \emptyset$, then $u_j\in \hat{N}_{t^*_1}(u_i)$, and we know that $|\hat{N}_{t^*_1}(u_i)|\le (2/\alpha)\log n$ (since $u_i\in V_l^\prime$). Therefore, for a random variable $x_i$, corresponding to node $u_i$, there are at most $(2/
\alpha)\log n$ variables $x_j$ such that $\textrm{Cov}(x_i,x_j)\ne 0$. Now, by utilizing the fact that $\textrm{Cov}(x_i,x_j)\le 1$ for two Bernoulli random variables $x_i, x_j$, we can conclude that $$\sum_{1\le i\ne j\le k}\textrm{Cov}(x_i,x_j)\le \frac{2k\log n}{\alpha}.$$ 

Furthermore, Since $x_i$ is a Bernoulli random variable, $\textrm{Var}[x_i]\le 1$. Thus, $\sum_{i=1}^{k}\textrm{Var}[x_i]\le k$. Overall, we get
\begin{equation}
\label{equation 2_var}
\textrm{Var}[X_l]\le k+\frac{2k\log n}{\alpha}\le \frac{(2\log n+1)n}{\alpha}.
\end{equation}
By applying Chebyshev's inequality and using Equations~(\ref{equation 1-expectation}) and~(\ref{equation 2_var}), we get
\begin{align*}
\Pr[X_l\ge \frac{\alpha\epsilon n}{2}]&\le \Pr[X_l\ge \mathbb{E}[X_l]+\frac{\alpha\epsilon n}{4}]\\ &\le \frac{\textrm{Var}[X_l]}{\alpha^2\epsilon^2 n^2/16}\\ &\le \frac{16(2\log n+1)n}{\alpha^3 \epsilon^2 n^2}\\ & \le 1/8\sqrt{n}.
\end{align*}~\qed
\begin{lemma}
\label{delta-half-weak}
For a graph $G=(V,E)$ and an ($\alpha$,$\epsilon$)-weak attacker, if half of the nodes are of degree at least $d^*_1:=(6/\alpha\epsilon^2)\log (12/\mu\alpha\epsilon)$, then for any $t\in\mathbb{N}$ the attacker cannot win in $t$ rounds w.p. at least $1-\mu$.
\end{lemma}
\proof Let $V_h$ be a node set of size $n/2$ whose nodes are all of degree at least $d^*_1$ and define $\overline{V}_h:=V\setminus V_h$. We prove that after one round w.p. at least $1-\mu$ the number of white nodes is more than $n/2$ (that is, the attacker cannot win for any number of rounds). In the following, we provide three statements whose combination yields the above claim.

Firstly, let $X_1:=|\overline{V}_h\cap W_0|$. Then, we have $\mathbb{E}[X_1]=(1/2+\epsilon)\alpha n/2=\alpha n/4+\alpha\epsilon n/2$. Very similar to the proof of Lemma~\ref{w_0-size-weak}, by applying the Chernoff bound (Theorem~\ref{Chernoff}) we can prove that $X_1\ge \alpha n/4+\alpha \epsilon n/4$ w.p. at least $1-\exp(-\Theta(\epsilon^2\alpha n))\ge 1-1/3\sqrt{n}\ge 1-\mu/3$ (where we used our assumptions that $\alpha, \epsilon$ are constant and $\mu\ge 1/\sqrt{n}$).

Furthermore, we define $X_2:=|V_h\cap B_0|$. Then, we have $\mathbb{E}[X_2]=(1/2-\epsilon)\alpha n/2=\alpha n/4-\alpha\epsilon n/2$. Again, by applying the Chernoff bound (Theorem~\ref{Chernoff}) we can show that $X_2\le \alpha n/4-\alpha \epsilon n/4$ w.p. at least $1-\exp(-\Theta(\epsilon^2 \alpha n))\ge 1-\mu/3$.

Finally, let $X_3:=|V_h\setminus (R_0\cup W_1)|$. Consider an arbitrary node $v\in V_h$. Let $Y_W$ and $Y_B$ respectively denote the number of white and black neighbors of $v$ in $R_0$. Suppose that $u_1,\cdots, u_d$ are the neighbors of $v$, where $d=\deg(v)\ge d^*_1$. We define Bernoulli random variable $y_i$ to be 1 if and only if $u_i$ is in $W_0$. Then, we have $\Pr[y_i=1]=(1/2+\epsilon)\alpha$ and $Y_W=\sum_{i=1}^{d}y_i$. Thus, we get $\mathbb{E}[Y_W]=(1/2+\epsilon)\alpha d$. By applying the Chernoff bound (Theorem~\ref{Chernoff}), we have
\begin{align*}
\Pr[Y_W\le \alpha d/2]& \le \Pr[Y_W\le (1-\epsilon)(1/2+\epsilon)\alpha d]\\ &= \Pr[Y_{W}\le (1-\epsilon)\mathbb{E}[Y_W]]\\ &\le \exp(-\epsilon^2(1/2+\epsilon)\alpha d/2)\\ &\le \exp(-\epsilon^2\alpha d^*_1/4)\\ &\le\alpha \epsilon \mu /12.
\end{align*}
Following a similar argument, one can show that $\Pr[Y_B\ge \alpha d/2]\le \alpha \epsilon \mu/12$. Therefore, we have $Y_W\le Y_B$ w.p. at most $\alpha\epsilon \mu/6$. Note that if $Y_W>Y_B$ and $v\notin R_0$, then we have $v\in W_1$. This implies that 
\[
\Pr[v\notin (R_0\cup W_1)]\le (1-\alpha)\alpha\epsilon \mu/6.
\]
Hence, $\mathbb{E}[X_3]\le (1-\alpha)\alpha \epsilon \mu n/12$. Now, by applying Markov's inequality (Theorem~\ref{Markov}), we get
\[
\Pr[X_3\ge \frac{\alpha\epsilon n}{4}]\le \frac{\mathbb{E}[X_3]}{\alpha\epsilon n/4}\le \frac{4(1-\alpha)\alpha\epsilon\mu n}{12\alpha \epsilon n}\le \mu/3.
\]

It remains to argue that the combination of the above three statements yields our claim. We proved that $X_2\le \alpha n/4-\alpha \epsilon n/4$ w.p. at least $1-\mu/3$ and $X_3< \alpha\epsilon n/4$ w.p. at least $1-\mu/3$. Recall that $X_2=|V_h\cap B_0|$ and $X_3=|V_h\setminus (R_0\cup W_1)|$. Thus, more than $n/2-(\alpha n/4-\alpha \epsilon n/4+\alpha \epsilon n/4)=n/2-\alpha n/4$ nodes in $V_h$ are white after one round w.p. at least $1-2\mu/3$. Furthermore, $X_1=|\overline{V}_h\cap W_0|\ge \alpha n/4+\alpha \epsilon n/4$ w.p. at least $1-\mu/3$. Overall, we can conclude that after one round there are at least $(n/2-\alpha n/4)+(\alpha n/4+\alpha\epsilon n/4)=n/2+\alpha\epsilon n/4$ white nodes, that is, the attacker cannot win after any number of rounds, w.p. at least $1-\mu$. \qed

\begin{customthm}{\ref{Delta-weak}}
For a graph $G=(V,E)$ and an ($\alpha$,$\epsilon$)- weak attacker, if $\Delta\le Cn/((\log n)^{C^{\prime}}\log (4/\mu))$ for some suitable constants $C_{\alpha, \epsilon}, C_{\alpha,\epsilon}^{\prime}>0$ or if half of the nodes are of degree at least $d^*_1:=(6/\alpha\epsilon^2)\log (12/\alpha\epsilon \mu)$, then for any $t\in\mathbb{N}$ the attacker cannot win in $t$ rounds w.p. at least $1-\mu$.
\end{customthm}
\proof Based on Lemma~\ref{delta-half-weak}, if half of the nodes are of degree at least $d_1^*$, then the attacker cannot win for any number of rounds w.p. at least $1-\mu$. Thus, we only need to prove that if $\Delta$ is smaller than the given bound, then after $t$ rounds, for any $t\in \mathbb{N}$, more than half of the colored nodes are white w.p. at least $1-\mu$.

Assume $t':=\min (t,t^*_1)$, where $t^*_1=(2/\alpha)\log (4/\alpha\epsilon)$. Let $V_l:=\{v\in V: \deg(v)\le d^*_2\}$, for $d_2^*=8\log n/(\alpha\epsilon^2)$, be the set of \emph{low-degree} nodes and $V_h:=V\setminus V_l$ be the set of \emph{high-degree} nodes. Then, we define $w_{t'}^l:=|V_l\cap (\hat{W}_{t'}\setminus W_0)|$ (analogously, $b_{t'}^{l}:=|V_l\cap (\hat{B}_{t'}\setminus B_0)|$) to be the number of non-seed low-degree nodes which are colored white (resp. black) in the first $t'$ rounds. Similarly, we define $w_{t'}^h$ and $b_{t'}^h$ for non-seed high-degree nodes.

Based on Lemma~\ref{w_0-size-weak}, we have $w_0\ge b_0+\alpha\epsilon n$ w.p. at least $1-\exp(-\Theta(\epsilon^2\alpha n))\ge 1-1/4\sqrt{n}\ge1-\mu/4$. Furthermore based on Lemma~\ref{high-degree-weak}, we know that $b_{t'}^h=0$ w.p. at least $1-2/n\ge 1-1/8\sqrt{n}\ge1-\mu/8$. We prove that $w_{t'}^l> b_{t'}^l-\alpha\epsilon n/2$ w.p. at least $1-\mu/4$. Note that $\hat{w}_{t'}=w_0+w_{t'}^l+w_{t'}^h$ and $\hat{b}_{t'}=b_0+b_{t'}^l+b_{t'}^h$. Therefore, the combination of the above three statements implies that $\hat{w}_{t'}> \hat{b}_{t'}+\alpha\epsilon n/2$ w.p. at least $1-3\mu/4$. If $t'=t$, then we are done. Otherwise, we have $t>t'=t^*_1$. Based on Lemma~\ref{time-small-weak}, we know that at most $\alpha\epsilon n/2$ nodes get colored after the $t^*_1$-th round w.p. at least $1-\mu/4$. In the worst case scenario, they all are colored black. Since $\hat{w}_{t_1^*}> \hat{b}_{t_1^*}+\alpha\epsilon n/2$ w.p. at least $1-3\mu/4$, we can conclude that $\hat{w}_t>\hat{b}_t$ w.p. at least $1-\mu$.

It remains to prove that $w_{t'}^l>b_{t'}^l-\alpha\epsilon n/2$ w.p. at least $1-\mu/4$. Let us consider a \emph{modified} version of our process. This is the same as the original process except that when it is time for a non-seed high-degree node to be colored, it adopts white color, regardless of the color of its neighbors. Note that based on Lemma~\ref{high-degree-weak}, we know that all non-seed high-degree nodes are colored white in the original process w.p. at least $1-\mu/8$. That is, the original process is the same as the modified one w.p. at least $1-\mu/8$. Therefore, if we show that $w_{t'}^l> b_{t'}^l-\alpha\epsilon n/2$ in the modified process w.p. at least $1-\mu/8$, we can conclude that $w_{t'}^l> b_{t'}^l-\alpha\epsilon n/2$ in the original process w.p. at least $1-\mu/4$. (We are basically conditioning on the event that all non-seed high-degree nodes are colored white.) 

Now, we prove that $w_{t'}^l> b_{t'}^l-\alpha\epsilon n/2$ in the modified process w.p. at least $1-\mu/8$. Recall that $r_0$ denotes the number of seed nodes and define $l:=|V_l\cap \cup_{j=1}^{t'}R_j|$ to be the number of low-degree nodes which are colored during rounds 1 to $t'$. (We should emphasize that both $r_0$ and $l$ are random variables. Thus to be precise, in our calculations below we need to condition on $r_0=r_0^{\prime}$ or $l=l'$ for some $r_0^{\prime},l'\in \mathbb{N}$. However for the sake of simplicity, we abuse the notation and treat $r_0$ and $l$ as fixed values and show that our arguments hold regardless of the value of $r_0$ and $l$.) Our goal is to prove that $w_{t'}^l> l/2-\alpha\epsilon n/4$ (because this implies that $w_{t'}^l> b_{t'}^l-\alpha\epsilon n/2$ by using $l=b_{t'}^l+w_{t'}^l$) w.p. at least $1-\mu/8$.

Let us label the seed nodes from $v_1$ to $v_{r_0}$. Corresponding to each node $v_i$, for $1\le i\le r_0$, we define the probability space $(\Omega_i,\Pr_i)$. Then, $w_{t'}^{l}:\Omega\rightarrow \mathbb{R}$ is defined over $\Omega$, where $(\Omega, \Pr)$ is the product of discrete probability spaces $(\Omega_i, \Pr_i)$ for $1\le i\le r_0$. Recall that the effect of the $i$-th coordinate is said to be at most $c_i$ if for all $\omega, \omega'\in \Omega$, which differ only in the $i$-th coordinate we have $|w_{t'}^{l}(\omega)-w_{t'}^{l}(\omega')|\le c_i$. We say a node $u$ is \emph{affected} by $v_i$ if switching the color of $v_i$ can result in $u$ adopting a different color. Hence, $c_i$ is upper-bounded by the number of nodes in $V_l\cap \cup_{j=1}^{t'}R_j$ which are affected by $v_i$. We observe that each node $v_i$ has at most $\Delta$ low-degree neighbors in $R_1$ which might be affected by $v_i$. Each of these nodes has at most $d^*_2$ low-degree neighbors in $R_2$ which might be affected by $v_i$. Applying this argument for $t'-2$ more times, we can conclude that $c_i\le 2\Delta (d^*_2)^{t'-1}\le \Delta (d^*_2)^{t'}$. Note that we do not take the high-degree nodes into account since in the modified process they are deterministically colored white. (We should mention that our goal is to apply Azuma's inequality (Theorem~\ref{Azuma}) and for that we need to upper-bound $\sum_{i=1}^{r_0}c_i^2$. A trivial bound is $r_0\Delta^2 (d^*_2)^{2t'}$, but this is too loose for our purpose here. We give a tighter bound by using a more careful counting argument.)

Let $u$ be a low-degree node in $R_j$ for some $1\le j\le t'$. Node $u$ is affected by a node $v_i$ if at least one of its low-degree neighbors in $R_{j-1}$ is affected by $v_i$. (Again, recall that high-degree nodes choose their color deterministically, it is why we only account for low-degree ones.) Since $u$ is a low-degree node, the number of such neighbors is at most $d^*_2$. Similarly, each of these neighbors is affected by $v_i$ if at least one of its low-degree neighbors in $R_{j-2}$ is affected by $v_i$. Repeating this argument, we reach some low-degree nodes in $R_1$, which are affected by $v_i$ if they are adjacent to it. Hence, we can conclude that each low-degree node $u$ in $\cup_{j=1}^{t'}R_j$ is affected by at most $2(d^*_2)^{t'}$ nodes in $R_0$. This implies that $\sum_{i=1}^{r_0}c_i\le 2l(d^*_2)^{t'}\le 2n(d^*_2)^{t'}$.

So far we showed that $c_i\le\Delta(d^*_2)^{t'}$ and $\sum_{i=1}^{r_0}c_i\le 2n(d^*_2)^{t'}$. Thus, we have
\[
\sum_{i=1}^{r_0}c_i^2\le \Delta(d^*_2)^{t'}\sum_{i=1}^{r_0}c_i\le 2n\Delta (d^*_2)^{2t'}.
\]
Recall that
\[
\Delta\le \frac{Cn}{(\log n)^{C^{\prime}}\log (4/\mu)},\ d^*_2=\frac{8\log n}{\alpha\epsilon^2}, \textrm{and}\ t'\le \frac{2}{\alpha}\log \frac{4}{\alpha \epsilon}.
\]
Therefore, for a suitable choice of $C_{\alpha, \epsilon}$ and $C_{\alpha,\epsilon}^{\prime}$ we get
\begin{align*}
\Delta\le \frac{\alpha^2\epsilon^2 n}{64\log (4/\mu)(d^*_2)^{2t'}}.
\end{align*}
Thus, we have $\sum_{i=1}^{r_0}c_i^2\le \alpha^2\epsilon^2 n^2/(32\log (4/\mu))$. Furthermore, since each node in $V_l \cap \cup_{j=1}^{t'}R_j$ is white w.p. at least $1/2$, we have that $\mathbb{E}[w_{t'}^l]\ge l/2$. Now, applying Azuma's inequality (Theorem~\ref{Azuma}) yields 
\begin{align*}
\Pr[w_{t'}^l \le \frac{l}{2}-\frac{\alpha\epsilon n }{4}] &\le \exp\left(-\frac{\alpha^2\epsilon^2 n^2}{32\sum_{i=1}^{r_0}c_i^2}\right)\\ &\le \exp\left(-\log \left(4/\mu\right)\right)=\mu/4.\ \ \ \ \ \ \ \ \ \ \Box
\end{align*}

\section{Tightness of Theorem~\ref{Delta-weak}}
\label{appendix-weak-tightness}
Consider some arbitrary $0<\epsilon, \alpha, \mu<1/2$ and let $s=(\log_{(\frac{1}{2}-\epsilon)\alpha}\mu)-1$. Consider an $n$-node graph $G$ which includes a node set $S$ of size $s$ and each node in $V\setminus S$ is adjacent to all nodes in $S$.
For an ($\alpha$,$\epsilon$)-weak attacker, we have $S\subset B_0$ w.p. $(\alpha(1/2-\epsilon))^s>\mu$. We observe that if $S\subseteq B_0$, then all uncolored nodes adopt black color in the first round. Then, the attacker wins w.p. more than $\mu$. Note that all nodes in $G$ are of degree at least $s$ which is equal to $C^{\prime\prime}\log (1/\mu)$ for a suitable choice of $C^{\prime\prime}_{\alpha,\epsilon}>0$. Therefore, the bound $(6/\alpha\epsilon^2)\log (12/\alpha\epsilon \mu)$ in Theorem~\ref{Delta-weak} is tight in terms of $\mu$. However, we believe that the dependency on $\alpha$ and $\epsilon$ is not best possible.

Now, consider an $n$-node graph $G'$, which is the union of $s$ star graphs, each of size $n/s$. Again, an ($\alpha$,$\epsilon$)-weak attacker will select all the $s$ internal nodes to be seed nodes and color them black w.p. $(\alpha(1/2-\epsilon))^s>\mu$. If all the internal nodes are black initially, then all the uncolored nodes adopt black color in the next round. Therefore, the attacker wins w.p. more than $\mu$. Note that the maximum degree of $G'$ is equal to $(n/s)-1=C_{\alpha,\epsilon}^{\prime\prime\prime}n/\log(1/\mu)$ for some constant $C_{\alpha,\epsilon}^{\prime\prime\prime}>0$. Therefore, the upper bound on $\Delta$ in Theorem~\ref{Delta-weak} is tight in terms of $\mu$. The dependency on $n$ is also optimal, up to the term $(\log n)^{C^{\prime}}$. Recall from the proof of Theorem~\ref{Delta-weak} that $C^{\prime}=(4/\alpha)\log (4/\alpha\epsilon)$. We can replace $C^{\prime}$ with $\log C^{\prime}$ by applying a fairly simple idea. However, to entirely get rid of this poly-logarithmic term, one perhaps needs proof techniques beyond the ones used in this paper.
\section{Tightness of Theorem~\ref{stabilization-time-random}}
\label{appendix-thigtness-1}
We prove that the upper bound in Theorem~\ref{stabilization-time-random} is tight, up to a constant factor. For an arbitrary integer $\delta\ge 1$, let $C_n^{\delta}$ be the $\delta$-th power of a cycle $C_n$, which is a graph on the same node set, but two nodes are adjacent if their distance is between 1 and $\delta$ in $C_n$. (For example, $C_n^1=C_n$.) Assume that we select each node to be a seed node independently w.p. $\alpha$. Since $C_n^{\delta}$ is $2\delta$-regular, based on Theorem~\ref{stabilization-time-random} the stabilization time is in $\mathcal{O}((1/\alpha\delta)\log n)$ a.a.s. On the other hand, we prove that there is at least one node whose $t^*_3$-neighborhood does not include any seed node a.a.s. for $t^*_3:=(1/16\alpha\delta)\log_2 n-1$. This implies that this node will be colored after the $t_3^*$-th round, i.e., the process takes more than $t_3^*=\Omega((1/\alpha\delta)\log n)$ rounds to end. Thus, the upper bound given in Theorem~\ref{stabilization-time-random} is asymptotically tight.

We observe that for a node $u$ in $\mathcal{C}_n^{\delta}$, we have $|\hat{N}_{t^*_3}(u)|=2\delta t^*_3+1\le (1/8\alpha)\log_2 n$. Thus, we can select $\sqrt{n}$ nodes $u_1,\cdots, u_{\sqrt{n}}$ whose $t^*_3$-neighborhood is pair-wise disjoint. Now, we define Bernoulli random variable $x_i$ to be 1 if and only if there is no seed node in $\hat{N}_{t^*_3}(u_i)$. Then, we have
\[
\Pr[x_i=1]\ge (1-\alpha)^{\frac{\log_2 n}{8\alpha}}\ge 4^{\frac{-\log_2 n}{8}}=n^{-\frac{1}{4}}.
\]
Let $X:=\sum_{i=1}^{\sqrt{n}} x_i$. Since the $t^*_3$-neighborhood of $u_i$s is pair-wise disjoint, we get
\[
\Pr[X=0]\le \left(1-n^{-\frac{1}{4}}\right)^{\sqrt{n}}\le \exp\left(-n^{\frac{1}{4}}\right).
\]
Thus, a.a.s. there is a node $u_i$ whose $t^*_3$-neighborhood does not include any seed node.

\section{Proof of Theorem~\ref{stabilization-time-lower}}
\label{appendix-stabilization-time-lower}
\begin{customthm}{\ref{stabilization-time-lower}}
	For a connected graph $G=(V,E)$, if each node is a seed node independently w.p. $\alpha$, then the stabilization time is in $\Omega(\log_{\Delta}\log n^{\frac{1}{\alpha}})$ a.a.s.
\end{customthm}

\proof Let $t^*_4:=\log_{\Delta}\log_2 n^{\frac{1}{16\alpha}}$. We first want to show that there is a large set of nodes whose $t^*_4$-neighborhood is pair-wise disjoint. To prove that, consider the following recursive procedure. Suppose that initially all nodes are \emph{unmarked} and consider a set $U=\emptyset$. As long as there is an unmarked node $u$, add it to $U$ and mark all nodes in $\hat{N}_{2t^*_4}(u)$. By construction, the $t^*_4$-neighborhood of each pair of nodes in $U$ is disjoint. Furthermore, when the procedure ends, $U$ includes at least $\sqrt{n}$ nodes since we mark at most $2\Delta^{2t^*_4}=2\log_2^2 n^{\frac{1}{16\alpha}}$ (i.e., poly-logarithmically many) nodes every time we add a node to $U$ and we start with $n$ unmarked nodes. 

So far, we proved that there exist $\sqrt{n}$ nodes $u_1,\cdots, u_{\sqrt{n}}$ whose $t^*_4$-neighborhood is pair-wise disjoint. Now, we define Bernoulli random variable $x_i$ to be 1 if and only if $\hat{N}_{t^*_4}(u_i)\cap R_0= \emptyset$. Since $|\hat{N}_{t^*_4}(u_i)|\le 2\Delta^{t^*_4}=(1/8\alpha)\log_2n$, we have
\[
\Pr[x_i=1]\ge (1-\alpha)^{\frac{\log_2 n}{8\alpha}}\ge 4^{-\frac{\log_2 n}{8}}=n^{-\frac{1}{4}}.
\]
Let $X:=\sum_{i=1}^{\sqrt{n}}x_i$. Since the $t^*_4$-neighborhood of $u_i$s is pair-wise disjoint, $\Pr[X=0]\le (1-n^{-1/4})^{\sqrt{n}}\le \exp(-n^{1/4})$. Thus, a.a.s. there is a node $u_i$ such that $\hat{N}_{t^*_4}(u_i)\cap R_0=\emptyset$, which implies that $u_i$ is colored after the $t^*_4$-th round. Hence, the stabilization time is at least $t^*_4=\Omega(\log_{\Delta}\log n^{\frac{1}{\alpha}})$. \qed

\section{Tightness of Theorem~\ref{stabilization-time-lower}}
\label{appendix-thigtness-2}
Consider an $n$-node complete $d$-ary tree $T=(V,E)$ for an arbitrary integer $d\ge 2$. We prove that the stabilization time is in $\mathcal{O}(\log_{d}\log n^{\frac{1}{\alpha}})$ a.a.s. when we select each node to be a seed node independently w.p. $\alpha>0$. This implies that the bound in Theorem~\ref{stabilization-time-lower} is asymptotically tight.

Let $L_i$ denote the set of nodes in depth $i$ and assume that $t^*_5:=\log_{d}\log n^{\frac{2}{\alpha}}+1$. Consider an arbitrary node $v\in V$. We claim that $|\hat{N}_{2t^*_5}(v)|\ge (2/\alpha)\log n$. If $v\in L_i$ for some $i\le\lfloor \log_{d}n\rfloor-t^*_5$, then it has at least $d^{t^*_5-1}= (2/\alpha)\log n$ nodes in its subtree which are in distance at most $t^*_5$ from $v$. Thus, we have $|\hat{N}_{2t^*_5}(v)|\ge |\hat{N}_{t^*_5}(v)|\ge (2/\alpha)\log n$. On the other hand, if $v\in L_i$ for some $i>\lfloor \log_{d}n\rfloor-t^*_5$, then it has a node $u$ in its $t^*_5$-neighborhood which is in $L_j$ for some $j\le\lfloor \log_{d}n\rfloor-t^*_5$. Since $|\hat{N}_{t^*_5}(u)|\ge (2/\alpha)\log n$ and $\hat{N}_{t^*_5}(u)\subseteq \hat{N}_{2t^*_5}(v)$, we have $|\hat{N}_{2t^*_5}(v)|\ge (2/\alpha)\log n$.

So far, we showed that $|\hat{N}_{2t^*_5}(v)|\ge (2/\alpha)\log n$ for any node $v$. Applying Lemma~\ref{logn-coin_flip} implies that a.a.s. each node has a seed node in its $2t^*_5$-neighborhood. Therefore, a.a.s. in at most $2t^*_5=\mathcal{O}(\log_{d}\log n^{\frac{1}{\alpha}})$ rounds, all nodes are colored and the process is over.

\section{Proof of Theorem~\ref{hardness-MI-problem}}
\label{appendix-hardness-MI-problem}
Before proving Theorem~\ref{hardness-MI-problem}, let us define the Minimum Influence problem, more formally.
\\
\\
\textsc{Minimum Influence (MI)}\\
\textit{Instance}: Graph $G=(V,E)$, and $b, w, t\in \mathbb{N}$.\\
\textit{Output}: Minimum expected number of white nodes after $t$ rounds (i.e., $\hat{w}_t$) if $b_0=b$ and $w_0=w$. 

\begin{customthm}{\ref{hardness-MI-problem}}
	Let $\mathcal{A}$ be a polynomial-time $\beta$-approximation algorithm for the MI problem. Then, $\beta>n^{1-\zeta}$ for any $\zeta>0$ unless P=NP.
\end{customthm}
\proof Assume that there is a polynomial-time $\beta$-approximation algorithm $\mathcal{A}$ which solves the MI problem, for $\beta=n^{1-\zeta}$ and some $\zeta>0$. Then, we prove that there exists a polynomial-time algorithm for the \textit{clique problem}, which is one of Karp's 21 NP-complete problems. In the clique problem, we are given a graph $G'$ and an integer $k$ as the input and the task is to determine whether $G'$ has a clique of size $k$ or not.

Let an $n'$-node graph $G'$ and an integer $k$ be the input of the clique problem. We will describe an $n$-node graph $G$, which can be constructed in polynomial time. Then, we prove that the solution of the MI problem on $G$ and for $b=k$, $w={k \choose 2}$, and $t=2$ is ${k \choose 2}$ if $G'$ has a clique of size $k$ and more than $n^{1-\zeta}{k\choose 2}$ if it does not. Hence, the output of algorithm $\mathcal{A}$ for $G$, $b=k$, $w={k\choose 2}$, and $t=2$ is between ${k\choose 2}$ and $n^{1-\zeta}{k\choose 2}$ if and only if $G'$ has a clique of size $k$. This implies that we have a polynomial-time algorithm for the clique problem, which is not possible unless P=NP.

Now, let us explain how we construct $G$, step by step. Consider a copy of $G'$ on the node set $V'=\{v_1,\cdots, v_{n'}\}$. Let $\{v_{i},v_{i'}\}$ be an edge in this graph, then we remove this edge and instead we add a new node $u$ and two edges $\{u,v_i\}$ and $\{u,v_{i'}\}$. Let us label these newly added nodes from $u_1$ to $u_{m'}$, where $m'$ is the number of edges in $G'$. Furthermore, corresponding to each node $v_i$ for $1\le i\le n'$ (analogously, node $u_j$ for $1\le j\le m'$) we add a node set $V_i$ (resp. $U_j$) of size $s=n'^{\frac{4}{\zeta}}$ and add a clique between nodes in $\{v_i\}\cup V_i$ (resp. $\{u_j\}\cup U_j$). To complete the construction of graph $G$, corresponding to each node $u_j$ for $1\le j\le m'$ we add a node $q_j$ and edge $\{u_j,q_j\}$. Therefore, nodes in $\{v_i\}\cup V_i$ induce a clique of size $s+1$ and nodes in $\{u_j,q_j\}\cup U_j$ induce a clique of size $s+1$ and a leaf attached to it. Overall, graph $G$ has $n=n'(s+1)+m'(s+2)\le n'^2s=n'^{2+\frac{4}{\zeta}}$ nodes.

Assume that $G'$ has a clique of size $k$. Then, we show that the MI problem for $b=k$, $w={k\choose 2}$, and $t=2$ on graph $G$ has a solution of size ${k\choose 2}$. Assume that a node set $Z\subseteq V'$ induces a clique of size $k$ in $G'$. By our construction, the node set $Z$ also exists in $G$, but each edge of the clique has been replaced with a gadget on nodes in $\{u_j,q_j\}\cup U_j$ for some $1\le j\le m'$. Let $U_Z$ and $Q_Z$ be the node sets which include all of these $u_j$s and $q_j$s respectively. We know that $|Q_Z|={k\choose 2}$ since $Z$ induces a clique in $G'$. Assume that we color all nodes in $Z$ black and all nodes in $Q_Z$ white. Thus, there are $k$ black nodes and ${k\choose 2}$ white nodes. We observe that each node $u_j\in U_Z$ is colored black in the next round since it has two black neighbors in $Z$ and only one white neighbor, namely $q_j$. Since $u_j$ is the only neighbor of $q_j$, the number of white nodes remains unchanged after any number of rounds.

It remains to prove that if $G'$ has no clique of size $k$, then the solution to the MI problem for $b=k$, $w={k\choose 2}$, and $t=2$ on $G$ is at least $s/2$. Recall that $n\le n'^{2+\frac{4}{\zeta}}$ and $s=n'^{\frac{4}{\zeta}}$. Thus, we have
\[
\frac{s}{2}\ge {n'\choose 2}n'^{\frac{4}{\zeta}-2}\ge {k
	\choose 2}n'^{(\frac{4}{\zeta}+2)(1-\zeta)}>{k\choose 2}n^{1-\zeta}.
\]
For the sake of contradiction, assume that there is an initial coloring of $G$ with $b=k$ black nodes and $w={k\choose 2}$ white nodes such that $\mathbb{E}[\hat{w}_2]< s/2$. Then, let us partition the nodes of $G$ into four different types of node sets. For $1\le i\le n'$, a set $\{v_i\}\cup V_i$ is a $v^b$-node set if $v_i$ is colored black initially and it is a $v^w$-node set otherwise. Furthermore for $1\le j\le m'$, a set $\{u_j,q_j\}\cup U_j$ is a $u^b$-node set if both neighbors of $u_j$ in $V'$ are black and it is a $u^w$-node set otherwise. Note that this way we partition the nodes of $G$ into $n'+m'$ node sets. 

For a node set $A$, let $b(A)$ and $w(A)$ denote the number of black and white nodes in $A$ in the initial coloring. We prove that
\begin{itemize}
\item $w(A)\le b(A)$ for a $v^w$-node set $A$
\item $w(A)\le b(A)-1$ for a $v^b$-node set $A$
\item $w(A)\le b(A)$ for a $u^w$-node set $A$
\item $w(A)\le b(A)+1$ for a $u^b$-node set $A$.
\end{itemize}
Let $g$ denote the number of white nodes minus the number of black ones in the initial coloring. We want to find an upper bound on $g$. We define $n_{v^b}$ and $n_{u^b}$ to be the number of $v^b$-node sets and $u^b$-node sets, respectively. The combination of the above four claims implies that $g\le n_{u^b}-n_{v^b}$. If $n_{v^b}=k$, then $n_{u^b}\le {k\choose 2}-1$. This is because $G'$ has no clique of size $k$ and thus there cannot be more than ${k\choose 2}-1$ $u_j$ nodes such that both of their neighbors in $V'$ are black. Hence, in this case $g<{k\choose 2}-k$. If $n_{v^b}=k'$ for some $k'<k$, then $g\le {k'\choose 2}-k'<{k\choose 2}-k$. (We used the fact that ${k'\choose 2}-k'<{k\choose 2}-k$ for $k'<k$.) Thus, we proved that $g<{k\choose 2}-k$. This contradicts our assumption that there are $k$ black nodes and ${k\choose 2}$ white nodes initially. Therefore, we can conclude that there is no initial coloring with $b_0=k$ and $w_0={k\choose 2}$ such that $\mathbb{E}[\hat{w}_2]<s/2$.

Now, we prove the above four claims regarding the relation of $w(A)$ and $b(A)$, one by one. First, consider a $v^w$-node set $A=\{v_i\}\cup V_i$. We must have $w(A)\le b(A)$ because otherwise all uncolored nodes in $A$ choose white in the next round. In that case, we will have more than $s/2$ white nodes since $|A|=s+1$.

Consider a $v^b$-node set $A=\{v_i\}\cup V_i$. If $w(A)\ge b(A)$, then the expected number of white nodes in $A$ after one round is at least $|A|/2=(s+1)/2>s/2$. Thus, we must have $w(A)\le b(A)-1$.

Consider a $u^w$-node set $A=\{u_j,q_j\}\cup U_j$. We prove that if $w(A)\ge b(A)+1$, then in expectation there are more than $s/2$ white nodes in two rounds (which implies that we must have $w(A)\le b(A)$). Obviously, we have $w(A\setminus \{q_j\})\ge b(A\setminus \{q_j\})$. If there is at least one colored node in $A\setminus \{q_j\}$, then in the next round at least $|A\setminus \{q_j\}|/2=(s+1)/2>s/2$ nodes will be white in expectation. Otherwise, all nodes in $A\setminus \{q_j\}$ are uncolored and $q_j$ is white. Note that since $A$ is a $u^w$-node set, at most one of the $u_j$'s neighbors in $V'$ is black. Thus, w.p. at least $1/2$, after one round $u_j$ becomes white and after another round all nodes in $A$ will be colored white. Thus, the expected number of white nodes after two rounds is at least $|A|/2>s/2$.

Finally, consider a $u^b$-node set $A$. If $w(A)\ge b(A)+2$, then $w(A\setminus \{q_j\})\ge b(A\setminus \{q_j\})+1$. In that case, all uncolored nodes in $A\setminus \{q_j\}$ become white in the next round; that is, there will be more than $s/2$ white nodes after one round. Thus, we must have $w(A)\le b(A)+1$. \qed

\end{document}